\documentclass[11pt,dvips]{article}
cm \voffset = -2truecm

\usepackage{graphicx}
\usepackage{amssymb}
\usepackage{latexsym}

\begin{document}

\setcounter{page}{1}
\renewcommand{\thefootnote}{\fnsymbol{footnote}}

\vspace{5mm}
\begin{center}

{\large \bf Quantum information and quantum computing: \\ an overview and some mathematical aspects\footnote{Paper from an invited talk at the 20th International Workshop on Computer Algebra in Scientific Computing (CASC, Lille, 2018). This paper does not appear in the proceedings of CASC 2018 because it was not ready in due time.}}

Maurice R. Kibler{\footnote{email: {\sf m.kibler@ipnl.in2p3.fr} }}

{\small   CNRS/IN2P3, Institut de Physique Nucl\'eaire, 69622 Villeurbanne, France \\ 
				  Facult\'e des Sciences et Technologies, Universit\'e Claude Bernard Lyon 1, 69622 Villeurbanne, France \\ 
          Universit\'e de Lyon, 69361 Lyon, France} 

\end{center}

\begin{abstract}
The aim of the present paper is twofold. First, to give the main ideas behind quantum computing and quantum information, a field based on quantum-mechanical phenomena. Therefore, a short review is devoted to (i) {\em quantum bits} or qubits (and more generally {\em qudits}), the analogues of the usual bits 0 and 1 of the classical information theory, and to (ii) two characteristics of quantum mechanics, namely, {\em linearity} (which manifests itself through the superposition of qubits and the action of unitary operators on qubits) and {\em entanglement} of certain multi-qubit states (a resource that is specific to quantum mechanics). Second, to focus on some mathematical problems related to the so-called {\em mutually unbiased bases} used in quantum computing and quantum information processing. In this direction, the construction of mutually unbiased bases is presented via two distinct approaches: one based on the group SU(2) and the other on Galois fields and Galois rings.  
\end{abstract}

\section{Introduction}

In the present days, there is a growing interest for the field of {\em quantum information} and {\em quantum computing}. Such a field emerged at the beginning of the 1980s when Feynman (and other scientists) asked the question: is it possible to simulate the behaviour of a quantum system by using a classical computer? Then, the question evolved towards how to use quantum systems to do computations. This led to the idea of a quantum computer based on quantum physics with the hope to solve problems that would be intractable or difficult to solve with a classical computer. A fact in favour of a quantum computer is the law by Moore according to which the size of electronic and spintronic devices for a classical computer should approach 10 nm in 2020, the scale where quantum effects become important. The field of quantum information and quantum computing is at the crossroads of experimental and theoretical quantum physical sciences (physics and chemistry), discrete mathematics and informatics with the aim of building a quantum computer. We note in passing that physics, mathematics, informatics and engineering have already greatly benefited from the enormous amount of works achieved along the line of quantum information and quantum computing.

The unit of classical information is the bit (possible values 0 and 1). In a quantum computer, classical bits (0 and 1) are replaced by {\em quantum bits} or {\em qubits} (that interpolate in some sense between 0 and 1). The most general qubit is a normalized vector $| \psi \rangle$ in the two-dimensional Hilbert space $\mathbb{C}^2$
\begin{eqnarray}
| \psi \rangle = a | 0 \rangle + b | 1 \rangle, 
\quad |a|^2 + |b|^2 = 1, 
\quad a \in \mathbb{C}, 
\quad b \in \mathbb{C}
\label{quantumbit}
\end{eqnarray}  
where $| 0 \rangle$ and $| 1 \rangle$ are the elements of an orthonormal basis in $\mathbb{C}^2$. The result of a measurement of $| \psi \rangle$ is not deterministic since it gives $| 0 \rangle$ or $| 1 \rangle$ with the probability $|a|^2$ or $|b|^2$, respectively. The consideration of $N$ qubits leads to work in the $2^N$-dimensional Hilbert space $\mathbb{C}^{2^N}$. Note that the notion of qubit, corresponding to $\mathbb{C}^2$, is a particular case of the one of {\em qudit}, corresponding to $\mathbb{C}^d$ ($d$ not necessarily in the form $2^N$). A system of $N$ qudits is associated with the Hilbert space $\mathbb{C}^{d^N}$. In this connection, the techniques developed for finite-dimensional Hilbert spaces are of paramount importance in quantum computation and quantum computing. 

From a formal point of view, a quantum computer is a system producing qubits, the state of which can be controlled and manipulated via unitary transformations. These transformations correspond to the product of elementary unitary operators called {\em quantum gates} (the analogues of the logic gates of a classical computer) acting on one, two or more qubits. Measurement of the qubits out-coming from a {\em quantum circuit} of quantum gates yields the result of a (quantum) computation. In other words, a realization of quantum information processing can be performed by preparing a quantum system in a quantum state, then submitting this state to unitary transformations and, finally, reading the outcome from a measurement.  

The two basic characteristics of quantum mechanics used in a quantum computer are {\it linearity} (principle of superposition of quantum states) and 
{\it entanglement}. The superposition principle gives resources: the quantum computer can be in several states at the same time. This leads to a massive quantum parallelism with a speed up of computations (for $N$ qubits, $2^N$ computations can be achieved in parallel through the use of {\it quantum algorithms}). Entanglement, i.e.~the fact that certain quantum systems made of two or more sub-systems behave as an indissociable entity, is at the root of quantum computing and {\it quantum teleportation}. In quantum mechanics, each measurement on a quantum system perturbs the system and the superposition principle makes impossible to duplicate a quantum state ({\it no-cloning theorem}). The two latter points and the use of the so-called 
{\em mutually unbiased bases} (MUBs), to be defined in Section 3, are the basic ingredients of quantum cryptography (illustrated by the BB84 protocol, the first protocol of quantum cryptography). 

The aim of this paper is to present to a community of computer engineers and mathematicians the basic grounds of quantum information and quantum computing as well as some mathematical aspects and related open problems. 

This paper is organized as follows. Section 2 deals with the general framework of quantum information and quantum computing (i.e.~information and computing based on quantum physics): some of the concepts and ideas evoked above are further described. In Section 3, we address some mathematical aspects of quantum information; in particular, we review some of the methods for constructing mutually unbiased bases (more precisely, methods based on the group SU(2) and on Galois rings and Galois fields). Sections 2 and 3 are mainly based on Refs.~\cite{01Nielsen} and \cite{02Kibler}, respectively. References \cite{Weyl}-\cite{Rao} constitute an incomplete list (in chronological order) of original works of relevance for an in-depth study of Sections 2 and 3. Finally, the reader will find in Ref.~\cite{Trifa55} some calculations with the help of the Python language illustrating the derivation of mutually unbiased bases according to the methods described in Section 3. 

\section{The general framework of quantum information and quantum computing} 

\subsection{Quantum mechanics in a few words}

Classical physics does not apply in the microscopic world. It is not appropriate for describing, explaining and predicting physical and chemical phenomena at the atomic and sub-atomic level. The convenient theory for quantum systems (i.e.~molecules, atoms, nuclei and elementary particles) is quantum mechanics, an extension of the old quantum theory mainly due to Planck, Einstein, Bohr and Sommerfeld (the word {\it quantum} comes from the fact that the energy exchanges between light and matter occur in a quantized form). Quantum mechanics, which is often used in conjunction with some other theories like relativity and quantum field theory, can be presented in two equivalent ways: {\it wave mechanics} initiated by de Broglie and 
Schr\"odinger and {\it matrix mechanics} pioneered by Heisenberg, Born and Jordan. It is not our purpose to list in detail the postulates of quantum mechanics. We shall restrict ourselves with four aspects of the Copenhagen interpretation which are indispensable in quantum information and quantum computing. 

$\bullet$ In both presentations of quantum mechanics, the state of a closed quantum system is described by a vector (in matrix mechanics) or a wave function (in wave mechanics), noted $| \psi \rangle$ in both cases, belonging to a finite or infinite Hilbert space $\cal H$. 

$\bullet$ In quantum information and quantum computing, the space $\cal H$ is finite-dimensional (isomorphic to $\mathbb{C}^2$ for qubits or 
$\mathbb{C}^d$ for qudits) and the (normalized) vector $| \psi \rangle$, defined up to a phase factor, can be the result (arising from an evolution or transformation of a vector $| \psi' \rangle$)
$$ 
| \psi \rangle = U | \psi' \rangle
$$
of the action of a unitary operator $U$ (or quantum gate) on $| \psi' \rangle$. (We are not concerned here with 
dynamical systems for which the time evolution of $\psi$ in the wave picture is given by the Schr\"odinger equation, in the non-relativistic case, or the Dirac equation, in the relativistic case, two linear equations). 

$\bullet$ In quantum information and quantum computing, $| \psi \rangle$ is given by a linear combination of the eigenvectors of an observable in the matrix formulation. An observable ${\cal A}$ is associated with a measurable physical quantity (energy, position, impulsion, spin, etc.). It is represented by a self-adjoint operator $A$ acting on the space $\cal H$. The possible outcomes of a measurement of an observable ${\cal A}$ are the real eigenvalues of the operator $A$. Measurement in quantum mechanics exhibits a probabilistic nature. More precisely, if (in the case of the finite-dimensional Hilbert space ${\cal H} = \mathbb{C}^d$) 
\begin{eqnarray} 
| \psi \rangle = \sum_{n=0}^{d-1} c_n | \varphi_n \rangle, \quad c_n \in \mathbb{C}
\label{psi en phi_n}
\end{eqnarray} 
where the $\varphi_n$ given by 
\begin{eqnarray*} 
A | \varphi_i \rangle = \lambda_i 
  | \varphi_i \rangle, \quad i = 0, 1, \cdots, d-1
\end{eqnarray*} 
are the orthormalized eigenvectors of $A$, then a measurement of ${\cal A}$ will give $\lambda_k$ with the probability 
\begin{eqnarray*} 
| c_k |^2 = | \langle \varphi_k | \psi \rangle |^2
\end{eqnarray*} 
where $\langle \varphi_k | \psi \rangle$ stands for the inner product of $| \psi \rangle$ by $| \varphi_k \rangle$ (we suppose that the spectrum of $A$ is non-degenerate). Hence, before measurement, the quantum system is in several states being a linear combination of the states $| \varphi_n \rangle$ and, after measurement, the quantum system is in a well-defined state $| \varphi_k \rangle$. Measurement leads to a reduction of the wave packet or 
wave function collapse. In terms of measurement of qudits, what precedes can be formulated as follows. Let $| \psi \rangle$ as given by 
Eq.~(\ref{psi en phi_n}) be a qudit describing a quantum system before measurement. A measurement of $| \psi \rangle$ in a basis 
$\{ \varphi_i \rangle : i = 0, 1, \cdots, d-1 \}$ of $\mathbb{C}^d$ yields the state 
\begin{eqnarray*}
\frac{\langle \varphi_i | \psi \rangle}{\sqrt{\langle \psi | \varphi_i \rangle \langle \varphi_i | \psi \rangle}} | \varphi_i \rangle = 
\frac{\langle \varphi_i | \psi \rangle}{| \langle \varphi_i | \psi \rangle |} | \varphi_i \rangle 
\end{eqnarray*}
with the probability 
$$
p(i) = | \langle \psi | \varphi_i \rangle |^2
$$
Observe that the factor ${\langle \varphi_i | \psi \rangle}{| \langle \varphi_i | \psi \rangle |^{-1}}$ is a simple phase factor without importance. By way of example, in the case of $\mathbb{C}^2$, measurement of the qubit $| \psi \rangle = a | 0 \rangle + b | 1 \rangle$ in the basis 
$\{ | 0 \rangle , | 1 \rangle \}$ of $\mathbb{C}^2$ yields $| 0 \rangle$ or $| 1 \rangle$ (up to unimportant phase factors) with the probabilities 
$|a|^2$ or $|b|^2$, respectively. 

$\bullet$ A postulate of quantum mechanics of considerable interest in quantum information and quantum computing concerns the description of a system composed of several sub-systems. The state vector for the system is build from tensors products of the state vectors of the various sub-systems. This may lead to entangled vector states for the composite system. Entanglement constitutes another important resource for quantum information and quantum computing besides the linearity and the non deterministic nature of quantum mechanics. As an example, suppose we have a system of qubits made of two two-level sub-systems. The Hilbert space for the system is ${\cal H} = {\mathbb{C}}^4 \sim {\mathbb{C}}^2 \otimes {\mathbb{C}}^2$, where the first and second $\mathbb{C}^2$ corresponds to the first and second sub-systems, respectively. By tensor product, we can take 
\begin{eqnarray*} 
\{ | 0 \rangle_1 \otimes | 0 \rangle_2, \ 
   | 0 \rangle_1 \otimes | 1 \rangle_2, \ 
	 | 1 \rangle_1 \otimes | 0 \rangle_2, \ 
	 | 1 \rangle_1 \otimes | 1 \rangle_2    \}
\end{eqnarray*}
as a basis for $\mathbb{C}^4$, where the indices 1 and 2 refer to the first and the second qubits, respectively. Two kinds of states can be considered in $\mathbb{C}^4$, namely separable or non entangled states as 
\begin{eqnarray*} 
| \psi_s \rangle = | 0 \rangle_1 \otimes \frac{1}{2} (| 0 \rangle_2 + \sqrt{3} | 1 \rangle_2)
\end{eqnarray*} 
and non separable or entangled states as 
\begin{eqnarray*} 
| \psi_{ns} \rangle = \frac{1}{\sqrt{2}} (| 0 \rangle_1 \otimes | 1 \rangle_2 +  
                                          | 1 \rangle_1 \otimes | 0 \rangle_2) 
\end{eqnarray*}
For the non entangled state $| \psi_s \rangle$, measurement of the qubit 1 yields $| 0 \rangle_1$ with the probability 1 while measurement of the qubit 2 leads either to $| 0 \rangle_2$ with the probability $\frac{1}{4}$ or to $| 1 \rangle_2$ with the probability $\frac{3}{4}$; therefore, the result of a measurement for one qubit does not depend on the result of a measurement for the other qubit. The situation turns out to be entirely different for the entangled state $| \psi_{ns} \rangle$: a measurement of the first qubit leads either to $| 0 \rangle_1$ with the probability 
$\frac{1}{2}$ or to $| 1 \rangle_1$ with the probability $\frac{1}{2}$; once one of the two results has been obtained, we immediately know what would be the result if we perform a measurement on the second qubit; it is thus unnecessary to make a measurement on the second qubit and this may be sum up as follows: 
\begin{eqnarray*} 
{\rm result \ of \ a \ measurement \ of \ qubit \ 1} & \Rightarrow & {\rm state \ of \ qubit \ 2 \ (without \ measurement)} \\ 
| 0 \rangle_1 & \Rightarrow & | 1 \rangle_2 \\
| 1 \rangle_1 & \Rightarrow & | 0 \rangle_2 
\end{eqnarray*} 
and conversely 
\begin{eqnarray*} 
{\rm result \ of \ a \ measurement \ of \ qubit \ 2} & \Rightarrow & {\rm state \ of \ qubit \ 1 \ (without \ measurement)} \\ 
| 1 \rangle_2 & \Rightarrow & | 0 \rangle_1 \\
| 0 \rangle_2 & \Rightarrow & | 1 \rangle_1 
\end{eqnarray*} 
Entanglement may also occur for more than two qubits. For entangled states, there are strong correlations between the results of measurements of the qubits. This effect is essential for quantum information and quantum computing. 

Unfortunately, ``Something is rotten in the state of Denmark'' (where the Copenhagen interpretation developed). In fact, entanglement is also an inconvenience: entanglement of qubits with their environment leads to errors. This is known as the effect of decoherence an important drawback for the building of a quantum computer. One way to fight against errors due to decoherence and other effects of noise is to develop {\it quantum error-correcting codes}. 

\subsection{Qubits and qudits}

\subsubsection{Qubits}

Let 
$$
B_2 = \{ | 0 \rangle , | 1 \rangle \} 
$$
be an orthonormal basis called the computational basis of the Hilbert space $\mathbb{C}^2$. Any normalized (to unity) vector $| \psi \rangle$, see Eq.~(\ref{quantumbit}), in $\mathbb{C}^2$ is called a quantum bit or qubit. From the quantum mechanical point of view, a qubit describes a state of a two-level quantum system. In the absence of measurement (and decoherence), the state $| \psi \rangle$ is a superposition of $| 0 \rangle$ and $| 1 \rangle$. A measurement of the state $| \psi \rangle$ yields either $| 0 \rangle$ (with the probability $|a|^2$) or $| 1 \rangle$ (with the probability $|b|^2$). Therefore, the superposition of the states $| 0 \rangle$ and $| 1 \rangle$ is lost after the measurement. In matrix form, we take 
$$
| 0 \rangle = \pmatrix{
1 \cr
0 \cr
}, \quad 
| 1 \rangle = \pmatrix{
0 \cr
1 \cr
}, \quad | \psi \rangle = \pmatrix{
a \cr
b \cr
}
$$  
From a group-theoretical point of view, $| 0 \rangle$ and $| 1 \rangle$ can be considered as the basis vectors for the fundamental irreducible representation $\left( \frac{1}{2} \right)$ of SU$(2)$, in the chain ${\rm SU}(2) \supset {\rm U}(1)$, with
\begin{eqnarray}
| 0 \rangle = | \frac{1}{2},  \frac{1}{2} \rangle, \quad 
| 1 \rangle = | \frac{1}{2}, -\frac{1}{2} \rangle
\label{NMR}
\end{eqnarray} 
in the notations of quantum angular momentum theory. 

The state $| \psi \rangle$ can be associated with a point $(x,y,z,t)$ of the sphere $S^3$ in $\mathbb{R}^4$ according to
\begin{eqnarray*} 
\mathbb{C}^2                  \rightarrow S^3 \ : \ 
a | 0 \rangle + b | 1 \rangle \mapsto     (x,y,z,t) 
\end{eqnarray*} 
with $a = x + {\rm i} y$ and $b = z + {\rm i} t$. In fact, the point $(x,y,z,t)$ can be visualized as a point $(1, \theta, \varphi)$ of the sphere $S^2$ in $\mathbb{R}^3$, referred to as the Bloch sphere, since $\psi$ can be re-written as 
\begin{eqnarray} 
\psi = \cos \frac{\theta}{2} | 0 \rangle + {\rm e}^{{\rm i} \varphi} 
       \sin \frac{\theta}{2} | 1 \rangle, \quad 0 \leq \theta \leq \pi, \quad 0 \leq \varphi < 2 \pi
\label{qubitenthetaphi}
\end{eqnarray} 
up to a global multiplicative phase factor. The application 
\begin{eqnarray*} 
S^3       \rightarrow S^2 \ : \ 
(x,y,z,t) \mapsto     (1, \theta, \varphi) 
\end{eqnarray*} 
corresponds to the first Hopf fibration $S^3 \stackrel{S^1}{\longrightarrow} S^2$ of compact fibre $S^1$. Any qubit as given by 
Eq.~(\ref{qubitenthetaphi}) can be represented by a point on the Bloch sphere. Table~\ref{Blochsphere} gives the correspondence between some remarkable qubits $| \psi \rangle$ and points on the Bloch sphere. Any unitary transformation acting on a qubit $| \psi \rangle$ corresponds to a rotation around an axis passing through the centre of the Bloch sphere. 

\begin{table} 
\begin{center}
\begin{tabular}{||c||c|c|c|c|c|c||}
\hline
$| \psi \rangle$ & $| 0 \rangle$ & 
                   $| 1 \rangle$ & 
									 $\frac{1}{\sqrt{2}} (| 0 \rangle + | 1 \rangle)$ & 
									 $\frac{1}{\sqrt{2}} (| 0 \rangle - | 1 \rangle)$ & 
									 $\frac{1}{\sqrt{2}} (| 0 \rangle + {\rm i} | 1 \rangle)$ & 
									 $\frac{1}{\sqrt{2}} (| 0 \rangle - {\rm i} | 1 \rangle)$
\\
\hline
$(\xi, \eta, \zeta)$ & $(0, 0, 1)$ & $(0, 0, -1)$ & $(1, 0, 0)$ & $(-1, 0, 0)$ & $(0, 1, 0)$ & $(0, -1, 0)$ 
\\
\hline
\end{tabular}
\caption{Correspondence between qubits $| \psi \rangle = \cos \frac{\theta}{2} | 0 \rangle + {\rm e}^{{\rm i} \varphi} 
                                                         \sin \frac{\theta}{2} | 1 \rangle$ and points
($\xi = \sin \theta \cos \varphi$, $\eta = \sin \theta \sin \varphi$, $\zeta = \cos \theta$) of the Bloch sphere $S^2$ in $\mathbb{R}^3$}
\label{Blochsphere}
\end{center}
\end{table}

Note that the sets
\begin{eqnarray}
B_0 = \left\lbrace \frac{| 0 \rangle + | 1 \rangle}        {\sqrt{2}}, \ 
                   \frac{| 0 \rangle - | 1 \rangle}        {\sqrt{2}}  \right\rbrace, \quad 
B_1 = \left\lbrace \frac{| 0 \rangle + {\rm i} | 1 \rangle}{\sqrt{2}}, \ 
                   \frac{| 0 \rangle - {\rm i} | 1 \rangle}{\sqrt{2}}  \right\rbrace, \quad
B_2 = \{ | 0 \rangle , | 1 \rangle \} 
\label{B0B1B2}
\end{eqnarray}
appearing in Table~\ref{Blochsphere} are three orthonormal bases of the space $\mathbb{C}^2$. In addition, the vectors in $B_0$, $B_1$ and $B_2$ are eigenvectors of the Pauli matrices $\sigma_1$, $\sigma_2$ and $\sigma_3$ (defined in Eq.~(\ref{Paulimatrices}) below), respectively. The bases $B_0$, $B_1$ and $B_2$ constitute the simplest example of the so-called MUBs to be studied in Section \ref{MUBs}. 

\subsubsection{Qudits}

The generalisation from the two-dimensional Hilbert space $\mathbb{C}^2$ to the $d$-dimensional Hilbert space $\mathbb{C}^d$ ($d > 2$) is immediate. Given an orthonormal basis (called the computational basis)
\begin{eqnarray}
B_d = \{ | n \rangle : n = 0, 1, \cdots, d-1 \}
\label{computationalbasis}
\end{eqnarray}
of $\mathbb{C}^d$, any normalized vector 
$$
| \psi \rangle = \sum_{n = 0}^{d-1} c_n | n \rangle, \quad \sum_{n = 0}^{d-1} |c_n|^2 = 1, \quad c_i \in \mathbb{C}, \quad i = 0, 1, \cdots, d-1
$$
is called a qudit. From the point of view of quantum mechanics, the states $| n \rangle$ can be realized as generalized angular momentum states with
\begin{eqnarray}
| n \rangle = | j , m \rangle, \quad n = j - m, \quad d = 2j+1 
\label{relationnjm}
\end{eqnarray}
where for fixed $j$, the index $m$ takes the values $-j, -j + 1, \cdots, j$. This yields the correspondence 
          \begin{eqnarray*}
| 0   \rangle = | j , j   \rangle, \quad
| 1   \rangle = | j , j-1 \rangle, \quad 
\cdots, \quad
| d-1 \rangle = | j , -j  \rangle
          \end{eqnarray*}
between qudits and angular momentum states. (Let us recall that the {\em angular momentum state} $| j , m \rangle$ is a common eigenstate of the square $J^2$ of a generalized angular momentum and of the $z$-component $J_z$ of the angular momentum.) Therefore, $| \psi \rangle$ can be re-written
$$
| \psi \rangle = \sum_{m = -j}^{j} d_{j - m} | j,m \rangle
$$
in the angular momentum basis $\{ | j,m \rangle : m = -j, -j + 1, \cdots, j\}$. For instance, a qutrit $| \psi \rangle$ can be written 
$$
| \psi \rangle = c_0 | 0 \rangle + c_1 | 1 \rangle + c_2 | 2 \rangle  
$$
in the ternary basis $\{ | 0 \rangle, | 1 \rangle, | 2 \rangle \}$ or 
$$
| \psi \rangle = d_2 | 1,-1 \rangle + 
                 d_1 | 1, 0 \rangle + 
								 d_0 | 1, 1 \rangle 
$$
in the balanced basis $\{ | 1,-1 \rangle, | 1,0 \rangle, | 1,1 \rangle \}$ associated with the angular momentum $j = 1$. 

\subsubsection{Qudits with $d = 2^N$}

In the case where $d = 2^N$, the corresponding qudits can be obtained from tensor products. For example, for $d = 4$ a basis of 
$\mathbb{C}^4 \sim \mathbb{C}^2 \otimes \mathbb{C}^2$ is 
\begin{eqnarray*}
& | 0 \rangle \otimes | 0 \rangle = \pmatrix{
1 \cr
0 \cr
} \otimes \pmatrix{
1 \cr
0 \cr
} = \pmatrix{
1 \cr
0 \cr
0 \cr
0 \cr
}, \quad 
| 0 \rangle \otimes | 1 \rangle = \pmatrix{
1 \cr
0 \cr
} \otimes \pmatrix{
0 \cr
1 \cr
} = \pmatrix{
0 \cr
1 \cr
0 \cr
0 \cr
} & \\ 
& | 1 \rangle \otimes | 0 \rangle = \pmatrix{
0 \cr
1 \cr
} \otimes \pmatrix{
1 \cr
0 \cr
} = \pmatrix{
0 \cr
0 \cr
1 \cr
0 \cr
}, \quad
| 1 \rangle \otimes | 1 \rangle = \pmatrix{
0 \cr
1 \cr
} \otimes \pmatrix{
0 \cr
1 \cr
} = \pmatrix{
0 \cr
0 \cr
0 \cr
1 \cr
} &
\end{eqnarray*}
Then, the most general quartit $| \psi \rangle$ is made of the superposition of tensor products of two qubits. In detail, we have 
\begin{eqnarray*}
| \psi \rangle = a | 0 \rangle \otimes | 0 \rangle + 
                 b | 0 \rangle \otimes | 1 \rangle + 
								 c | 1 \rangle \otimes | 0 \rangle + 
								 d | 1 \rangle \otimes | 1 \rangle 
\end{eqnarray*}
where $a, b, c, d \in \mathbb{C}$ (usually, in $| i \rangle \otimes | j \rangle$ the state $| i \rangle$ refers to the first qubit and $| j \rangle$ to the second). 

It is interesting to remark that the vectors $| \psi \rangle$ for $d = 2$, $2^2$ and $2^3$ are associated with the Hopf fibrations 
$S^3    \stackrel{S^1}{\longrightarrow} S^2$ (connected to complex numbers), 
$S^7    \stackrel{S^3}{\longrightarrow} S^4$ (connected to quaternions) and 
$S^{15} \stackrel{S^7}{\longrightarrow} S^8$ (connected to octonions). Entanglement for $d = 2^2$ and $2^3$ can be discussed in terms of fibrations on spheres \cite{Mosseri}. In the same vein, we may ask the question of the interest for entanglement of Cayley-Dickson algebras for $d = 2^N$ with $N > 3$ and of fibrations on hyperboloids \cite{Lambert}. 

\subsection{Physical realizations of qubits}

According to R.~Landauer, information is physical so that qubits are realised by quantum systems, more specifically by two-level quantum systems, the qubits $| 0 \rangle$ and $| 1 \rangle$ corresponding to two different (energy) levels. We shall not be concerned here with the physical realization of qubits (and qudits). It is enough to say that any two-level quantum system may be considered as a qubit. Therefore, qubits can be carried out by nuclear spins, trapped ions, neutral atoms and Bose-Einstein condensates, two different polarizations of a photon, and Josephson tunnel  nanojunctions. For instance, in nuclear magnetic resonance, the nuclear spins of an atom in an organic molecule can be aligned (giving the state $| 0 \rangle$) or anti-aligned (giving the state $| 1 \rangle$) with an applied constant magnetic field; in generalized angular momentum terminology, we have the quantum states given by Eq.~(\ref{NMR}) and corresponding to the spin $j = \frac{1}{2}$. Similarly, for an ion cooled and trapped by electric fields in a cavity, qubits can be implemented as electronic states (ground state for $| 0 \rangle$, excited state for $| 1 \rangle$). Vibrational states can also be used for realizing qubits (zero-phonon state for $| 0 \rangle$, one-phonon state for $| 1 \rangle$). 

\subsection{Entanglement}

\subsubsection{Generalities}

Entanglement occurs only in quantum physics. It has no analogue in classical physics. The notion of entanglement goes back to the famous paper by Einstein, Poldosky and Rosen. In quantum physics, two (or more than two) particles are said to be entangled if the quantum state of each particle depends of the quantum state(s) of the other(s) or cannot be described independently of the quantum state(s) of the other(s). In other words, there exist correlations between the physical properties of a system of entangled particles. More generally, two entangled sub-systems $S_1$ and $S_2$ are not independent so that the global system $\{S_1, S_2\}$ must be considered as a whole even after separation by an arbitrary distance. Then, a measurement made on one sub-system gives an information on the other (without measurement on the other sub-system). On the contrary, for a non entangled system consisting of two sub-systems, a measurement on one sub-system does not give in general an information on the other sub-system. 

As an example, let us consider a system consisting of two particles, system having a total spin equal to 0. If the spin of one particle is measured to be 
$\frac{1}{2}$ on a certain axis, then we know (without any measurement) that the spin on the other particle on the same axis is $-\frac{1}{2}$ because  
$$
0 = \frac{1}{2} -\frac{1}{2}
$$  
The two particles are not independent, even after separation. They still behave like an indivisible system of spin 0. 

Entanglement contradicts the principle of locality. There is non locality in the sense that what happens in some place depends of what happens in another place. Indeed, quantum mechanics is a non local, non deterministic and linear physical theory.  

\subsubsection{Entanglement of qubits} \label{entanglement}

In quantum information, the notion of entanglement occurs for multi-qubit systems. Let us consider a two-qubit system. There are two possibilities.

$\bullet$ The system is non entangled (or separable); it is then described by a state $ |\psi_s \rangle \in \mathbb{C}^2 \otimes \mathbb{C}^2 $ such that 
$$
|\psi_s \rangle = (a | 0 \rangle + b | 1 \rangle) \otimes (c | 0 \rangle + d | 1 \rangle)
$$
which can be re-written as 
$$
|\psi_s \rangle = ac | 0 \rangle \otimes | 0 \rangle + ad | 0 \rangle \otimes | 1 \rangle + bc | 1 \rangle \otimes | 0 \rangle + bd | 1 \rangle \otimes | 1 \rangle
$$
where $a | 0 \rangle + b | 1 \rangle$ and $c | 0 \rangle + d | 1 \rangle$ refer to the first and second qubit, respectively. 

$\bullet$ The system is entangled (or non separable); it is then described by a state $ |\psi_{ns} \rangle \in \mathbb{C}^4$ such that 
$$
|\psi_s \rangle = A | 0 \rangle \otimes | 0 \rangle + B | 0 \rangle \otimes | 1 \rangle + C | 1 \rangle \otimes | 0 \rangle + D | 1 \rangle \otimes | 1 \rangle
$$
cannot be written as the tensor product of two qubits in $\mathbb{C}^2$.  

It is clear that a necessary and sufficient condition for an arbitrary two-qubit state
$$
  |\psi \rangle = \alpha | 0 \rangle \otimes | 0 \rangle + \beta  | 0 \rangle \otimes | 1 \rangle 
                + \gamma | 1 \rangle \otimes | 0 \rangle + \delta | 1 \rangle \otimes | 1 \rangle
$$ 
of $\mathbb{C}^4$ to be non entangled is 
$$
\alpha \delta - \beta \gamma = 0 
$$
Therefore, if $\alpha \delta - \beta \gamma \not= 0$, then the state is entangled. The degree of entanglement of an arbitrary normalized two-qubit state 
$| \psi \rangle$ is characterized by the concurrence defined by 
\begin{eqnarray}
C = | \alpha \delta - \beta \gamma |, \quad 0 \leq C \leq \frac{1}{2}
\label{concurrence} 
\end{eqnarray}
Non entangled states correspond to $C = 0$, maximally entangled states to $C = \frac{1}{2}$. (A maximally entangled state is such that the density operator for each qubit is half the identity operator; it corresponds to a maximum value of the entropy.) Equation (\ref{concurrence}) can be straightforwardly generalized to the case 
$$
| \psi \rangle = \sum_{i = 0}^{d-1} \sum_{j = 0}^{d-1} a_{ij} | i \rangle \otimes | j \rangle
$$
of two qudits for which the concurrence $C$ is defined as 
$$
C = \det (a_{ij}), \quad  0 \leq C \leq \frac{1}{\sqrt{d^d}}
$$
in agreement with Eq.~(\ref{concurrence}) for $d = 2$. 

{\bf Example 1.} Let us consider the four states ($\oplus$ stands for the addition modulo 2)
$$
| \beta_{xy} \rangle = \frac{1}{\sqrt{2}} [| 0 \rangle   \otimes | y \rangle   + (-1)^x | 1 \rangle   \otimes | y \oplus 1 \rangle], \quad x,y = 0,1
$$    
called Bell states (in reference to the work on the so-called Bell inequalities) or EPR pairs 
(in reference to the paper by Einstein, Poldosky and Rosen). As a particular case
$$
| \beta_{01} \rangle = \frac{1}{\sqrt{2}} (| 0 \rangle_1 \otimes | 1 \rangle_2 +        | 1 \rangle_1 \otimes | 0 \rangle_2)
$$    
where the first qubit (qubit 1) and the second one (qubit 2) are clearly emphasized in order to avoid confusion. The result of a measurement of the qubit 1 gives

$\bullet$ either $| 0 \rangle_1$ (with the probability $\frac{1}{2}$) so that the qubit 2 is {\it a priori} (without measurement) in the state  $| 1 \rangle_2$

$\bullet$ or     $| 1 \rangle_1$ (with the probability $\frac{1}{2}$) so that the qubit 2 is {\it a priori} (without measurement) in the state  $| 0 \rangle_2$

{\noindent but no measurement can lead to both qubits 1 and 2 in the same state ($| 0 \rangle$ or $| 1 \rangle$). The result of a measurement of the qubit 1 provides information on the qubit 2 and reciprocally.} It is then unnecessary to make a measurement of one qubit once the result of the measurement of the other is known. Similar conclusions can be obtained for the three other Bell states $\beta_{00}$, $\beta_{10}$ and $\beta_{11}$. The four Bell states are maximally entangled (they correspond to $C=\frac{1}{2}$). 

In passing note that 
$$
| \beta_{xy} \rangle = (-1)^{xy} [(\sigma_1)^y (\sigma_3)^x] \otimes \sigma_0 | \beta_{00} \rangle
$$  
where $\sigma_0$, $\sigma_1$ and $\sigma_3$ are three of the four Pauli matrices
\begin{eqnarray}
\sigma_0 = \pmatrix{
1 & 0  \cr
0 & 1  \cr
}, \quad 
\sigma_1 = \pmatrix{
0 & 1  \cr
1 & 0  \cr
}, \quad 
\sigma_3 = \pmatrix{
1 & 0   \cr
0 & -1  \cr
}, \quad 
\sigma_2 = {\rm i} \sigma_1 \sigma_3 = \pmatrix{
0       & -{\rm i}  \cr
{\rm i} &        0  \cr
}
\label{Paulimatrices}
\end{eqnarray}
Thus, any Bell state $| \beta_{xy} \rangle$ can be obtained from $| \beta_{00} \rangle$. 

{\bf Example 2.} Let us consider the separable state 
\begin{eqnarray*} 
| \psi \rangle = (a | 0 \rangle + b | 1 \rangle) \otimes \frac{1}{\sqrt{5}} (| 0 \rangle + 2 | 1 \rangle) 
               = \frac{1}{\sqrt{5}} (a | 0 \rangle \otimes | 0 \rangle + 2 a | 0 \rangle \otimes | 1 \rangle 
                + b | 1 \rangle \otimes | 0 \rangle + 2 b | 1 \rangle \otimes | 1 \rangle)
\end{eqnarray*}
tensor product of two normalized qubits. A measurement of the first qubit gives either $| 0 \rangle$ with the probability 
$|a|^2 = |\frac{a}{\sqrt{5}}|^2 + |\frac{2a}{\sqrt{5}}|^2$ or $| 1 \rangle$ with the probability 
$|b|^2 = |\frac{b}{\sqrt{5}}|^2 + |\frac{2b}{\sqrt{5}}|^2$ while a measurement of the second qubit gives either $| 0 \rangle$ with the probability 
$\frac{1}{5} = |\frac{a} {\sqrt{5}}|^2 + |\frac{b} {\sqrt{5}}|^2$ or $| 1 \rangle $ with the probability 
$\frac{4}{5} = |\frac{2a}{\sqrt{5}}|^2 + |\frac{2b}{\sqrt{5}}|^2$. Therefore, a measurement on one qubit does not provide information on the other qubit (the state $| \psi \rangle$ corresponds to $C=0$). 

It is important to realize that entanglement of qubits (as in Example 1) and more generally of qudits has no analogue for classical bits. To be clear, the bits in $00$ or $01$ or $10$ or $11$ are not correlated. This is not the case for the quantum bits in any of the Bell states $\beta_{xy}$. 

\subsection{Quantum gates}

\subsubsection{One-qubit gates}

In a classical computer, bits are handled with the help of logic gates (there exist seven basic logic gates: AND, OR, XOR, NOT, NAND, NOR, and XNOR). A quantum computer processes qubits arranged in registers. It is equipped with quantum gates which perform unitary transformations on qubits. Quantum gates can be represented by unitary matrices. Table~\ref{4quantumgates} gives some examples of quantum gates [G] for one-qubit systems together with their matrix representations $G$. The actions of the one-qubit gates of Table~\ref{4quantumgates} on the qubit $| x \rangle$ (with $x = 0$ or $1$) are given by 
\begin{eqnarray*}
& | x \rangle \rightarrow [{\rm I}] \rightarrow | x \rangle, 
\quad | x \rangle \rightarrow [{\rm NOT}] \rightarrow | x \oplus 1 \rangle & \\
& | x \rangle \rightarrow [{\rm S}_\theta] \rightarrow {\rm e}^{{\rm i} x \theta} | x \rangle, 
\quad | x \rangle \rightarrow [{\rm H}] \rightarrow \frac{1}{\sqrt{2}} (| 0 \rangle + (-1)^x | 1 \rangle) 
\equiv \frac{1}{\sqrt{2}} ( | x \oplus 1 \rangle + (-1)^x | x \rangle ) &
\end{eqnarray*}
(as an example, the quantum circuit $| x \rangle \rightarrow [{\rm S}_\theta] \rightarrow {\rm e}^{{\rm i} x \theta} | x \rangle$ is described by the action 
$S_\theta | x \rangle = {\rm e}^{{\rm i} x \theta} | x \rangle$). Therefore, by linearity 
\begin{eqnarray*}
& a | 0 \rangle + b | 1 \rangle \rightarrow [{\rm NOT}] \rightarrow b | 0 \rangle + a | 1 \rangle & \\
& a | 0 \rangle + b | 1 \rangle \rightarrow [{\rm S}_\theta] \rightarrow a | 0 \rangle + {\rm e}^{{\rm i} \theta} b | 1 \rangle & \\ 
& a | 0 \rangle + b | 1 \rangle \rightarrow [{\rm H}] \rightarrow \frac{1}{\sqrt{2}} (a + b) | 0 \rangle + \frac{1}{\sqrt{2}} (a - b) | 1 \rangle & \\
& a | 0 \rangle + b | 1 \rangle \rightarrow [{\rm H}] \rightarrow [{\rm H}] \rightarrow a | 0 \rangle + b | 1 \rangle &
\end{eqnarray*}
(the last circuit reflects that $H^2 = I$). Note that the most general qubit can be obtained from the sequence  
$[{\rm H}] \rightarrow [{\rm S}_{2\theta}] \rightarrow [{\rm H}] \rightarrow [{\rm S}_{\frac{\pi}{2} + \varphi}]$ of one-qubit gates since 
$$
| 0 \rangle \rightarrow [{\rm H}] \rightarrow [{\rm S}_{2\theta}] \rightarrow [{\rm H}] \rightarrow [{\rm S}_{\frac{\pi}{2} + \varphi}] \rightarrow 
\cos \theta | 0 \rangle + {\rm e}^{{\rm i} \varphi} \sin \theta | 1 \rangle
$$
or
$$
S_{\frac{\pi}{2} + \varphi} H S_{2 \theta} H | 0 \rangle = \cos \theta | 0 \rangle + {\rm e}^{{\rm i} \varphi} \sin \theta | 1 \rangle
$$
up to the phase factor ${\rm e}^{{\rm i} \theta}$. 

\subsubsection{Multi-qubit gates} 

Quantum gates for two-qubit systems are important. For example, let us mention the controlled-NOT gate [${\rm C}_{\rm NOT}$] defined via 
$$
| x \rangle \otimes | y \rangle \rightarrow [{\rm C_{NOT}}] \rightarrow | x \rangle \otimes | y \oplus x \rangle 
$$
or in operator form
$$
C_{NOT} | x \rangle \otimes | y \rangle = | x \rangle \otimes | y \oplus x \rangle 
$$
where the first input qubit $| x \rangle$ and the second input qubit $| y \rangle$ are called control qubit and target qubit, respectively. Here, the corresponding quantum circuit has two inputs ($| x \rangle$ and $| y \rangle$) and two outputs ($| x \rangle$ and $| y \oplus x \rangle$). In matrix form, we have
the permutation matrix 
$$
C_{NOT} = \pmatrix{
1 & 0 & 0 & 0 \cr
0 & 1 & 0 & 0 \cr
0 & 0 & 0 & 1 \cr
0 & 0 & 1 & 0 \cr
}
$$ 
Clearly, $(C_{NOT})^2 = I$. Note that 
\begin{eqnarray}
C_{NOT} | x \rangle \otimes | 0 \rangle = | x \rangle \otimes | x \rangle 
\label{CNOTsurx0}
\end{eqnarray}
where $x = 0$ or $1$; however, this result does not mean that an arbitrary state $| \psi \rangle = a | 0 \rangle + b | 1 \rangle$ can be cloned by using the gate [${\rm C}_{\rm NOT}$] since we generally have (see Section \ref{nocloningth})
$$
C_{NOT} | \psi \rangle \otimes | 0 \rangle \not= | \psi \rangle \otimes | \psi \rangle
$$
to be compared with Eq.~(\ref{CNOTsurx0}). Note also that 
$$
| x \rangle \otimes | y \rangle \rightarrow [{\rm H} \otimes {\rm I}] \rightarrow [{\rm C_{NOT}}] \rightarrow | \beta_{xy} \rangle  
$$
or
$$
| \beta_{xy} \rangle = C_{NOT} (H \otimes I) | x \rangle \otimes| y \rangle
$$
that shows the interest of the gate [${\rm C_{\rm NOT}}$] for producing Bell states (i.e.~entangled states) from non entangled states. (By $[{\rm H} \otimes {\rm I}]$, we mean that the quantum gates [H] and [I] act on $| x \rangle$ and $| y \rangle$, respectively. Hence, $H \otimes I$ stands for the direct product of the matrices $H$ and $I$.) 

More generally, the quantum gate $[{\rm U}_{\rm f}]$ is defined through 
$$
| x \rangle \otimes | y \rangle \rightarrow [{\rm U}_{\rm f}] \rightarrow | x \rangle \otimes | y \oplus f(x) \rangle 
$$
or in an equivalent way
$$
U_f | x \rangle \otimes | y \rangle = | x \rangle \otimes | y \oplus f(x) \rangle 
$$
where $f$ stands for the function $f : \{ 0,1 \} \rightarrow \{ 0,1 \}$. Clearly, $(U_f)^2 = I$.

Another important two-qubit gate is the controlled phase gate $[{\rm CP}_\theta]$ such that 
$$
| x \rangle \otimes | y \rangle \rightarrow [{\rm CP}_\theta] \rightarrow | x \rangle \otimes {\rm e}^{{\rm i} x y \theta} | y \rangle 
$$
or 
$$
CP_\theta | x \rangle \otimes | y \rangle = | x \rangle \otimes {\rm e}^{{\rm i} x y \theta} | y \rangle 
$$
with
$$ 
CP_\theta = \pmatrix{
1 & 0 & 0 & 0 \cr
0 & 1 & 0 & 0 \cr
0 & 0 & 1 & 0 \cr
0 & 0 & 0 & {\rm e}^{{\rm i} \theta} \cr
}
$$ 
Note that 
$$
[{\rm C_{NOT}}] = [{\rm I} \otimes {\rm H}] \rightarrow [{\rm CP}_{\frac{\pi}{2}}] \rightarrow [{\rm CP}_{\frac{\pi}{2}}] \rightarrow [{\rm I} \otimes {\rm H}]
$$
so that the gate [${\rm C}_{\rm NOT}$] can be obtained from the gates [I $\otimes$ H] and [CP$_{\frac{\pi}{2}}$].

There exist other two-qubit gates. Moreover, use is also made of $n$-qubit gates ($n > 2$). The advantage of the quantum gates over the classical logic gates is that all the quantum gates are reversible or invertible due to the unitary property of the matrices representing quantum gates; this is not always the case for classical logic gates. 

The preceding examples are sufficient for illustrating how works the algorithm set up by Deutsch and Jozsa \cite{01Nielsen}.  

\begin{table} 
\begin{center}
\begin{tabular}{||c||c|c|c|c||}
\hline
gate [G] & identity gate [I] & not gate [NOT] & phase gate [S$_\theta$] & Hadamard gate [H] \\
\hline
matrix form $G$ & $I = \pmatrix{
1 & 0 \cr
0 & 1 \cr
}$ & 
$NOT = \pmatrix{
0 & 1 \cr
1 & 0 \cr
}$ & 
$S_\theta = \pmatrix{
1 & 0 \cr
0 & {\rm e}^{{\rm i} \theta} \cr
}$ & 
$H = \frac{1}{\sqrt{2}} \pmatrix{
1 & 1  \cr
1 & -1 \cr
}$  \\
\hline
\end{tabular}
\caption{Four basic quantum gates for one-qubit systems; the gates [I] and [NOT] also denoted [X] are associated with the Pauli matrices $\sigma_0$ or 
$I$ and $\sigma_1$ or $\sigma_x$, respectively; the two other Pauli matrices $\sigma_2$ or $\sigma_y$ and $\sigma_3$ or $\sigma_z$ define two further one-qubit gates denoted as [Y] and [Z], respectively} 
\label{4quantumgates}
\end{center}
\end{table}

\subsubsection{Quantum computing algorithms}

The Deutsch-Jozsa algorithm addresses the following problem: to find with only one measurement if the function 
$$
f : \{ 0,1 \}^{\otimes n} \rightarrow \{ 0,1 \}
$$
is constant or balanced 
($f$ is balanced means either $f(0) = 0$ and $f(1) = 1$ or $f(0) = 1$ and $f(1) = 0$; 
 $f$ is constant means $f(0) = f(1) = 0$ or $1$). The classical algorithm requires $2^{n-1} + 1$ evaluations of $f$ whereas only one measurement is necessary in order to get the answer. For $n = 1$, the proof based on the quantum circuit 
$[{\rm H} \otimes {\rm H}] \rightarrow [{\rm U}_{\rm f}] \rightarrow [{\rm H} \otimes {\rm I}]$ of two-qubit gates is as follows. It is easy to show that 
$$
| 0 \rangle \otimes | 1 \rangle \rightarrow [{\rm H} \otimes {\rm H}] \rightarrow [{\rm U}_{\rm f}] 
                    \rightarrow [{\rm H} \otimes {\rm I}] \rightarrow | x \rangle \otimes | y \rangle
$$
alternatively
$$
| x \rangle \otimes | y \rangle = (H \otimes I) U_f (H \otimes H) | 0 \rangle \otimes | 1 \rangle 
$$
where
\begin{eqnarray*}
| x \rangle \otimes | y \rangle &=& \pm | 0 \rangle \otimes \frac{1}{\sqrt{2}}(| 0 \rangle - | 1 \rangle) \ 
{\rm if} \ f \ {\rm is \ constant} \\
| x \rangle \otimes | y \rangle &=& \pm | 1 \rangle \otimes \frac{1}{\sqrt{2}}(| 0 \rangle - | 1 \rangle) \ 
{\rm if} \ f \ {\rm is \ balanced}
\end{eqnarray*} 
Then, the result of a single measurement of the first output qubit can be 
$$
\cases{
 | 0 \rangle \ \Rightarrow \ f \ {\rm is \ constant} \cr
 {\rm or} \cr
 | 1 \rangle \ \Rightarrow \ f \ {\rm is \ balanced} 
}
$$
Therefore, a single measurement (instead of two in the classical case) is sufficient for getting the answer. The Deutsch-Jozsa algorithm is of little interest. However, it shows the superiority of the quantum approach on the classical one (namely, only one measurement instead of $2^{n-1} + 1$ evaluations in the general case where $f : \{ 0,1 \}^{\otimes n} \rightarrow \{ 0,1 \}$). 

Let us briefly mention two other historical algorithms, viz, the Shor algorithm and the Grover algorithm \cite{01Nielsen}. The Shor algorithm concerns the search of the period of a periodic function and is used for the factorization of a composite integer into prime factors. It constitutes an alternative to the classical RSA code. The Grover algorithm makes it possible to find an item in an unstructured data basis consisting of $n$ entries; the quantum speed up for this algorithm is $n \rightarrow \sqrt{n}$ (O($n$) researches for the classical case and O($\sqrt{n}$) for the quantum case). The two preceding algorithms are based on the massive quantum parallelism. They formally show the superiority of a (still hypothetical) quantum computer on a classical one. The present evolution is towards quantum cryptography.
 
\subsection{No-cloning theorem} \label{nocloningth}

We may ask the question: does there exist a unitary operator (or quantum gate) $U$ such that 
\begin{eqnarray}
U | \psi \rangle \otimes | 0 \rangle = 
  | \psi \rangle \otimes 
	| \psi \rangle   
\label{nocloning1} 
\end{eqnarray}
where $| \psi \rangle = a | 0 \rangle + b | 1 \rangle$ is an arbitrary qubit. As a consequence of the linearity of quantum mechanics, the answer is no: it is not possible to clone an arbitrary qubit $| \psi \rangle$ \cite{05Wootters}. This result can be proved in the following way. Suppose that there exists $U$ such that Eq.~(\ref{nocloning1}) is true. Then, by linearity 
\begin{eqnarray}
U | \psi \rangle \otimes | 0 \rangle &=& U (a | 0 \rangle + b | 1 \rangle) \otimes | 0 \rangle                        \nonumber \\
                                     &=& U (a | 0 \rangle \otimes | 0 \rangle + b | 1 \rangle \otimes | 0 \rangle)    \nonumber \\
																		 &=&  a U | 0 \rangle \otimes | 0 \rangle + b U | 1 \rangle \otimes | 0 \rangle   \nonumber \\
																		 &=&  a   | 0 \rangle \otimes | 0 \rangle + b   | 1 \rangle \otimes | 1 \rangle   \label{nocloning2}
\end{eqnarray}
On another side, we have
\begin{eqnarray}
U | \psi \rangle \otimes | 0 \rangle &=& | \psi \rangle \otimes | \psi \rangle \nonumber \\
                                     &=& (a | 0 \rangle + b | 1 \rangle) \otimes (a | 0 \rangle + b | 1 \rangle) \nonumber \\
																		 &=& a^2 | 0 \rangle \otimes | 0 \rangle + ab (| 0 \rangle \otimes | 1 \rangle + | 1 \rangle \otimes | 0 \rangle) 
																		  +  b^2 | 1 \rangle \otimes | 1 \rangle \label{nocloning3}
\end{eqnarray}
Compatibility between Eqs.~(\ref{nocloning2}) and (\ref{nocloning3}) yields
$$
a^2 = a \ (\Rightarrow \ a = 0, 1), \quad 
b^2 = b \ (\Rightarrow \ b = 0, 1), \quad
a b = 0 \ (\Rightarrow \ a = 0 \ {\rm or} \ b = 0)
$$
The sole solutions are ($a = 1, \ b = 0$) and ($a = 0, \ b = 1$) in agreement with Eq.~(\ref{CNOTsurx0}). There are no solution in the general case. This proves the no-cloning theorem (a theorem that does not have an analogue in classical information). 

Another way to understand this result is to realize that in order to clone an arbitrary state $| \psi \rangle = a | 0 \rangle + b | 1 \rangle$ one must measure it so that one gets $| 0 \rangle$ or $| 1 \rangle$, two states that differ from $| \psi \rangle$ in general. 

\subsection{Quantum teleportation}

It is not possible to clone an arbitrary quantum state. However, it is feasible to teleporte it, i.e.~to transfer it from one place to another without an effective transportation. In other words, without a material transportation of a qubit, it is possible to transmit at distance the information contained in the qubit. We shall not deal here with some physical device making teleportation possible. We shall rather limit ourselves to the corresponding quantum algorithm \cite{10Bennett}. 

Suppose someone, Alice, wants to send a qubit $| \psi \rangle = a | 0 \rangle + b | 1 \rangle$ (for which she does not know $a$ and $b$) to somebody, Bob, by a quantum circuit and the possibility of using a classical communication channel. The only requirement for Bob and Alice is to dispose of and EPR pair 
$| \beta_{00} \rangle$, the first qubit of which belongs to Alice and the second one to Bob. Thus, the entry $| \varphi_0 \rangle$ of the quantum circuit is 
\begin{eqnarray*}
| \varphi_0 \rangle &=& | \psi \rangle \otimes | \beta_{00} \rangle \\
                    &=& (a | 0 \rangle_1 + b | 1 \rangle_1) \otimes \frac{1}{\sqrt{2}} 
										(| 0 \rangle_2 \otimes
										 | 0 \rangle_3 + 
										 | 1 \rangle_2 \otimes 
										 | 1 \rangle_3) \\
										&=& \frac{1}{\sqrt{2}} \big[ a | 0 \rangle_1 \otimes (| 0 \rangle_2 \otimes | 0 \rangle_3 + | 1 \rangle_2 \otimes | 1 \rangle_3) 
									                       	     + b | 1 \rangle_1 \otimes (| 0 \rangle_2 \otimes | 0 \rangle_3 + | 1 \rangle_2 \otimes |1 \rangle_3) \big]	\\                    &=& \frac{1}{\sqrt{2}} \big[ a(| 0 \rangle_1 \otimes | 0 \rangle_2 \otimes | 0 \rangle_3	+ 
																							     | 0 \rangle_1 \otimes | 1 \rangle_2 \otimes | 1 \rangle_3) + 
																								 b(| 1 \rangle_1 \otimes | 0 \rangle_2 \otimes | 0 \rangle_3  +
																								   | 1 \rangle_1 \otimes | 1 \rangle_2 \otimes | 1 \rangle_3) \big]																										\end{eqnarray*}
where qubits 1 and 2 refer to the Alice qubit and qubit 3 to the Bob qubit. Then, Alice sends her qubits to a controlled-NOT gate producing the state
\begin{eqnarray*}
| \varphi_1 \rangle &=& \frac{1}{\sqrt{2}} \big[ a(| 0 \rangle_1 \otimes | 0 \rangle_2 \otimes | 0 \rangle_3	+ 
																							     | 0 \rangle_1 \otimes | 1 \rangle_2 \otimes | 1 \rangle_3) + 
																								 b(| 1 \rangle_1 \otimes | 1 \rangle_2 \otimes | 0 \rangle_3  +
																								   | 1 \rangle_1 \otimes | 0 \rangle_2 \otimes | 1 \rangle_3) \big] \\
                    &=& \frac{1}{\sqrt{2}} \big[ a | 0 \rangle_1 \otimes ( | 0 \rangle_2 \otimes | 0 \rangle_3	+ | 1 \rangle_2 \otimes | 1 \rangle_3 ) + 
                                                 b | 1 \rangle_1 \otimes ( | 1 \rangle_2 \otimes | 0 \rangle_3  + | 0 \rangle_2 \otimes | 1 \rangle_3) \big]
\end{eqnarray*}
Next, qubit 1 goes to an Hadamard gate giving 
\begin{eqnarray*}
| \varphi_2 \rangle = \frac{1}{2} \big[ a ( | 0 \rangle_1 + | 1 \rangle_1) \otimes ( | 0 \rangle_2 \otimes | 0 \rangle_3	+ | 1 \rangle_2 \otimes | 1 \rangle_3 ) + 
                                        b ( | 0 \rangle_1 - | 1 \rangle_1) \otimes ( | 1 \rangle_2 \otimes | 0 \rangle_3  + | 0 \rangle_2 \otimes | 1 \rangle_3) \big]
\end{eqnarray*}
which can be re-arranged as
\begin{eqnarray*}
 | \varphi_2 \rangle &=& \frac{1}{2} \big[ (| 0 \rangle_1 \otimes | 0 \rangle_2) \otimes (a | 0 \rangle_3	+ b | 1 \rangle_3) + 
                                        (| 0 \rangle_1 \otimes | 1 \rangle_2) \otimes (a | 1 \rangle_3	+ b | 0 \rangle_3)  \\
																				& + & (| 1 \rangle_1 \otimes | 0 \rangle_2) \otimes (a | 0 \rangle_3	- b | 1 \rangle_3) + 
																				(| 1 \rangle_1 \otimes | 1 \rangle_2) \otimes (a | 1 \rangle_3	- b | 0 \rangle_3) \big] 
\end{eqnarray*}
Measurement of qubits 1 and 2 by Alice can give 
\begin{eqnarray*}
&| 0 \rangle_1 \otimes | 0 \rangle_2 \ {\rm with \ the \ probability} \ \frac{1}{4}, \quad
 | 0 \rangle_1 \otimes | 1 \rangle_2 \ {\rm with \ the \ probability} \ \frac{1}{4}& \\
&| 1 \rangle_1 \otimes | 0 \rangle_2 \ {\rm with \ the \ probability} \ \frac{1}{4}, \quad
 | 1 \rangle_1 \otimes | 1 \rangle_2 \ {\rm with \ the \ probability} \ \frac{1}{4}&
\end{eqnarray*}
Suppose Alice gets $| 0 \rangle_1 \otimes | 0 \rangle_2$. Then, she communicates this result to Bob by a classical channel (telephone or mail). Thus, Bob knows that 
$| \psi \rangle$ is $a | 0 \rangle_3	+ b | 1 \rangle_3$. 
Should Alice have got 
$| 0 \rangle_1 \otimes | 1 \rangle_2$ or 
$| 1 \rangle_1 \otimes | 0 \rangle_2$ or 
$| 1 \rangle_1 \otimes | 1 \rangle_2$, then Bob would obtain $a | 0 \rangle_3	+ b | 1 \rangle_3$ after the use of the gates [X] or [Z] or [T] with $T = ZX$ on the states $a | 1 \rangle_3 + b | 0 \rangle_3$ or  
       $a | 0 \rangle_3	- b | 1 \rangle_3$ or   
       $a | 1 \rangle_3	- b | 0 \rangle_3$, respectively. In all cases, the qubit $| \psi \rangle = a | 0 \rangle	+ b | 1 \rangle$ has been teleported. This proof shows that entanglement (via the EPR pair) plays a crucial role in teleportation.  

\section{Some mathematical aspects: mutually unbiased bases} \label{MUBs}

\subsection{Introducing MUBs}

\subsubsection{Generalities about MUBs}

Unitary operator bases in the Hilbert space $\mathbb{C}^{d}$ are of pivotal importance for quantum information and quantum computing as well as for quantum mechanics in general. The interest for unitary operator bases started with the seminal work by Schwinger \cite{03Schwinger}. In this connection, MUBs play a key role in quantum information and quantum computing. Two distinct orthonormal bases of $\mathbb{C}^d$ are said to be unbiased if and only if the modulus of the inner product of any vector of one basis with any vector of the other one is equal to 
$\frac{1}{\sqrt{d}}$ (see the detailed definition in Section \ref{def et prop MUBs}). 

MUBs proved to be useful in classical information theory (network communication protocols) \cite{11Calderbank,Heath}. They play an important role in quantum mechanics as for the discrete Wigner function \cite{Wootters1987,20Gibbons,Paz,Pittenger,Durt,39Bj,42Albouy}, for the solution of the Mean King problem \cite{14Englert,Aravind,Hayashi,Paz,Durt}, for the understanding of the Feynman path integral formalism \cite{Svetlichny,43Tolar} and potentially for studies of the Weyl-Heisenberg group in connection with quantum optics. MUBs are of central importance in quantum information theory as for instance in quantum state tomography (deciphering an unknown quantum state) \cite{Grassl,40Klimov,Rao}, quantum cryptography (secure quantum key exchange) \cite{06Bennett,18Cerf} and quantum teleportation \cite{10Bennett}. Along this line, measurements corresponding to MUBs are appropriate for an optimal determination of the density matrix of a quantum system and the use of MUBs ensure the maximum of security for quantum communication (especially in the BB84 quantum cryptography protocol). Let us also mention that MUBs are connected with the notion of maximal entanglement of quantum states a result of great importance for quantum computing. 

There exist numerous ways of constructing sets of MUBs (e.g., see \cite{Kibler,44Kibler,46Durt}). Most of them are based on discrete Fourier transform over Galois fields and Galois rings \cite{09Wootters,11Calderbank,19Klappenecker,20Gibbons,21Pittenger,24Archer,28Durt,51Ghiu,02Kibler}, discrete Wigner distribution \cite{Wootters1987,20Gibbons,Paz,Pittenger,39Bj}, generalized Pauli spin matrices \cite{15Bandyopadhyay,16Lawrence,Lawrence,21Pittenger}, mutually orthogonal Latin squares \cite{Hayashi,23Wocjan}, graph theory \cite{53Spengler}, finite and projective geometries \cite{33Klimov,41Appleby}, convex polytopes \cite{27Bengtsson}, complex projective 2-designs \cite{13Zauner,26Klappenecker,Zauner}, quantum angular momentum theory \cite{KiblerQTAM}, group theoretical methods \cite{17Chaturvedi,29Kibler,34Sulc,37Boykin}, discrete phase states \cite{48Daoud} and Hadamard matrices \cite{52Goyeneche}. In this section, from quantum theory of angular momentum theory (or, in mathematical terms, from the Lie algebra $A_1$ of the group SU$(2)$ or SU$(2 , \mathbb{C})$ or SL$(2 , \mathbb{C})$) we shall derive a formula for a complete set of MUBs in dimension $p$ with $p$ prime. Moreover, we shall construct complete sets of MUBs in dimension $p^m$ with $p$ prime and $m$ positive integer from the additive characters of the Galois field $\mathbb{GF}(p^m)$ for $p$ odd and of the Galois ring $\mathbb{GR}(2^2 , m)$ for $p=2$.

\subsubsection{Definition of MUBs} \label{def et prop MUBs}

\noindent $\blacktriangleright$~{\bf Definition.} Let $B_a$ and $B_b$ two distinct orthonormal bases 
\begin{eqnarray*}
  B_a = \{ | a \alpha \rangle \ : \ \alpha = 0, 1, \cdots, d-1 \},  \quad
  B_b = \{ | b \beta  \rangle \ : \ \beta  = 0, 1, \cdots, d-1 \} 
\end{eqnarray*}
of the Hilbert space $\mathbb{C}^d$. The bases $B_a$ and $B_b$ ($a \not= b$) are said to be unbiased if and only if 
\begin{eqnarray}
\forall \alpha \in \mathbb{Z}_d, \ 
\forall \beta  \in \mathbb{Z}_d  : | \langle  a \alpha |  b \beta \rangle | = \frac{1}{\sqrt{d}}
\label{definition des mubs}
\end{eqnarray}
where $\langle \ | \ \rangle$ denotes the inner product of $\mathbb{C}^d$ \cite{03Schwinger,04Ivanovic,07Wootters,Wootters1987}. In other words, the inner product $\langle a \alpha | b \beta \rangle$ has a modulus independent of $\alpha$ and $\beta$. The relation 
\begin{eqnarray*}
| \langle  a \alpha |  b \beta \rangle | = \delta_{a,b} \delta_{\alpha,\beta} + \frac{1}{\sqrt{d}} (1 - \delta_{a,b}) 
\end{eqnarray*}
makes it possible to describe both the cases $B_a = B_b$ and $B_a \not= B_b$.~$\blacktriangleleft$

As a typical example, the bases $B_0$, $B_1$ and $B_2$ of $\mathbb{C}^2$, see Eq.~(\ref{B0B1B2}), constitute a set of 3 MUBS whose basis vectors are specific qubits.  

\subsubsection{Well-known results about MUBs} \label{conjectures}

The main results concerning MUBs are \cite{04Ivanovic,09Wootters,13Zauner,23Wocjan,25Grassl}: 
\begin{enumerate}
	\item 
MUBs are stable under unitary or anti-unitary transformations. More precisely, if two unbiased bases undergo the same unitary or anti-unitary transformation, they remain mutually unbiased. 	
	\item
The number $N(d)$ of MUBs in $\mathbb{C}^d$ cannot exceed $d+1$. Thus 
$$
N(d) \leq d + 1
$$	
	\item 
The maximum number $d+1$ of MUBs is attained when $d$ is a power $p^m$ ($m \geq 1$) of a prime number $p$. Thus
$$
N(p^m) = p^m + 1
$$ 
	\item
When $d$ is a composite number, $N(d)$ is not known but it can be shown that 
$$
3 \leq N(d) \leq d + 1
$$  
As a more accurate result, for $d = \prod_i p_i^{m_i}$ with $p_i$ prime and $m_i$ positive integer, we have 
$$
{\rm min}(p_i^{m_i}) + 1 \leq N(d) \leq d + 1
$$  
\end{enumerate}

By way of illustration, let us mention the following cases. 
 \begin{itemize}
	 \item
In the particular composite case $d = 6 = 2 \times 3$, we have 
$$
3 \leq N(6) \leq 7 
$$
and it was conjectured that $N(6) = 3$. Indeed, in spite of an enormous amount of computational works, no more than three MUBs were found for $d=6$.
	\item 
For $d = 15 = 3 \times 5$ and $d = 21 = 3 \times 7$, there are at least four MUBs. 
	 \item 
For $d = 676 = 2^2 \times 13^2$, we have 
	$$
	2^2 + 1 = 5 \leq N(676) \leq 677
	$$
but it is known how to construct at least six MUBs. 
\end{itemize}

A set of $d+1$ MUBs in $\mathbb{C}^d$ is referred to as a complete set. Such sets exist for $d = p^m$ ($p$ prime, $m$ positive integer) and this result opens the way to establish a link between MUBs and Galois fields and/or Galois rings. 

For $d$ composite (different from a power of a prime), the question to know if there exist complete sets in dimension $d$, i.e.~to know if $N(d)$ can be equal to $d + 1$, is still an open problem (in 2018). Indeed, for $d$ different from a power of a prime, it was conjectured (SPR conjecture \cite{22Saniga}) that the problem of the existence of a set of $d+1$ MUBs in $\mathbb{C}^d$ is equivalent to the problem of whether there exist a projective plane of order $d$. As another conjecture for $d$ composite (different from a power of a prime), the problem of the existence of a set of $d+1$ MUBs in $\mathbb{C}^d$ is equivalent to the one of the existence of a decomposition of the Lie algebra of SU$(d)$ into $d+1$ Cartan subalgebras of dimension $d-1$.

\subsubsection{Interests of MUBs} 

MUBs are or relevance in advanced quantum mechanics. From a very general point of view, MUBs are closely connected to the principle of complementarity introduced by Bohr in the early days of quantum mechanics. This principle, quite familiar in terms of observables like position and momentum, tells that for two non-commuting observables, if we have a complete knowledge of one observable, then we have a total uncertainty of the other. Equation 
(\ref{definition des mubs}) indicates that the development in the basis $B_a$ of any vector of the basis $B_b$ is such that each vector of $B_a$ appears in the development with the probability $\frac{1}{d}$. This is especially interesting when translated in terms of measurements, the bases $B_a$ and $B_b$ corresponding to the (non-degenerate) eigenvectors of two non-commuting observables.

A significance of MUBs in terms of quantum measurements can be seen as follows. Let $A$ and $B$ be two non-degenerate (i.e.~with multiplicity-free eigenvalues) self-adjoint (or hermitian) operators associated with two observables ${\cal A}$ and ${\cal B}$ of a quantum system with the Hilbert space $\mathbb{C}^d$ of dimension $d$. Suppose that the eigenvectors of $A$ and $B$ yield two unbiased bases $B_a$ and $B_b$, respectively. When the quantum system is prepared in an eigenvector $| b \beta \rangle$ of the observable $B$, no information can be obtained from a measurement of the observable $A$. This result follows from the development in the basis $B_a$ of any vector of the basis $B_b$ 
$$
| b \beta  \rangle = \sum_{\alpha = 0}^{d-1}  | a \alpha \rangle    \langle  a \alpha |  b \beta \rangle
$$
which shows that the $d$ probabilities 
$$
| \langle  a \alpha |  b \beta \rangle |^2 = \frac{1}{d}, \quad \alpha, \beta = 0, 1, \cdots, d-1
$$
of obtaining any state vector $| a \alpha \rangle$ in a measurement of $A$ are equal. 

Indeed, the two operators $A$ and $B$ do not commute. The two corresponding observables ${\cal A}$ and ${\cal B}$ are said to be complementary (Bohr's principle of complementarity introduced in the early days of quantum mechanics): a precise knowledge of one of them implies a total uncertainty of the other (or, all possible results of measurements of the other one are equally probable). This can be made more explicit through the generalized Heisenberg uncertainty principle. Let $A$ and $B$ be two hermitian operators associated with two observables and $| \psi \rangle$ a vector of 
$\mathbb{C}^d$. The generalized Heisenberg uncertainty principle can be expressed as
\begin{eqnarray*}
\Delta A \Delta B \geq \frac{1}{2} | \langle \psi | [A,B]_- | \psi \rangle |
\end{eqnarray*}
where $[A,B]_- = AB - BA$ and $\Delta O$ stands for the standard deviation 
$$
\Delta O = \sqrt{ \langle \psi | O^2 | \psi \rangle - \langle \psi | O | \psi \rangle^2 }
$$
of the operator $O = A$ or $B$.\footnote{The most familiar example is for $d$ infinite. The position $A = x$ and the momentum 
$B = p_x$ (along the $x$ direction) of a particle are complementary observables. They satisfy the commutation relations $[x , p_x]_- = {\rm i} \hbar$, 
where $\hbar$ is the Planck constant. Hence, $\Delta x \Delta p_x \geq \frac{1}{2} \hbar$ so that more precise is $\Delta x$ more imprecise is 
$\Delta p_x$ and {\em vice versa}.} Therefore, if $A$ and $B$ correspond to observables generating MUBs, then a precise knowledge of $A$ yields a complete indeterminacy of $B$ and {\em vice versa}. 

Note that 
$$
d + 1 = \frac{d^2 - 1}{d - 1}
$$
is the number of different measurements to fully determine a quantum state for a quantum system in dimension $d$. (This follows from the fact that a 
$d \times d$ density matrix, that is to say an Hermitian matrix with a trace equal to 1, contains $d^2 - 1$ real parameters and each measurement gives 
$d-1$ real parameters.) Note also that $d^2 - 1$ and $d - 1$ are the number of generators and the rank of the special unitary group SU$(d)$ in $d$ dimensions, respectively, and that for $d = p$ (prime number) their ratio $p+1$ is the number of disjoint sets of $p-1$ commuting generators of SU$(p)$.

The rest of the paper is structured in the following way. In Section \ref{group theory}, we give a complete solution, based on a nonstandard approach to the Lie algebra of the group SU$(2)$ (equivalently, to the quantum theory of angular momentum), for the construction of MUBs in the case where $d = p$ is a prime number. Further developments are discussed in Section \ref{Weyl} in relation with Weyl pairs. Sections \ref{Galois field approach} and 
\ref{Galois ring approach} are concerned with the construction of MUBs from Galois fields (for $d = p^m$, a power of an odd prime number) and Galois rings (for $d = 2^m$, a power of the even prime number), respectively. (See Refs.~\cite{Vourdas,31Vourdas,32Vourdas} for the formalism of Galois quantum systems.) 

\subsection{Group-theoretical construction of MUBs} \label{group theory}

\subsubsection{Standard basis for SU$(2)$}

Equation (\ref{relationnjm}) shows that the vectors $| n \rangle$ (with $n = 0, 1, \cdots, d-1$) of the computational basis 
(\ref{computationalbasis}) can be viewed as the basis vectors $| j , m \rangle$ (with $m = j, j-1, \cdots, -j$) for the irreducible representation 
$(j)$ of SU$(2)$ in the chain ${\rm SU}(2) \supset {\rm U}(1)$. In the language of group theory (and quantum angular momentum theory), the vector 
$| j,m \rangle$ is a common eigenvector of the Casimir operator $J^2$ (the square of an angular momentum) and of a Cartan generator $J_z$ 
(the $z$ component of the angular momentum) of the Lie algebra $A_1$ of the group SU$(2)$. More precisely, we have the eigenvalue equations 
          \begin{eqnarray*}
          J^2 |j , m \rangle = j(j+1) |j , m \rangle, \quad 
          J_z |j , m \rangle = m      |j , m \rangle
          \end{eqnarray*}
with the orthonormality relations 
$$
\langle j , m | j , m' \rangle = \delta_{m , m'}, \quad m, m' = j, j-1, \cdots, -j
$$
In other words, the computational basis $B_d$ can be visualized as the basis
			\begin{eqnarray*}
B_{2j+1} = \{ | j , m \rangle \ : \ m = j, j-1, \cdots, -j \}
			\end{eqnarray*}
known as the standard basis for the irreducible representation $(j)$ of SU$(2)$ or the angular momentum basis corresponding to the angular momentum quantum number $j$, referred to as spin angular momentum for $j = \frac{1}{2}$. 

\subsubsection{Nonstandard bases for SU$(2)$} 

As far as the representation theory of SU$(2)$ is concerned, we can replace the complete set $\{ J^2, J_z \}$ by another complete set of two commuting operators. For instance, we may consider the set $\{ J^2, v_{a} \}$, where the unitary operator $v_{a}$ is defined by 
          \begin{eqnarray*}
v_{a} | j , m \rangle = \cases{ | j , -j \rangle \quad {\rm if} \quad m = j \cr \cr 
					                      \omega^{(j - m)a} | j , m + 1 \rangle \quad {\rm if} \quad m = j-1, j-2, \cdots, -j
					                    }	
          \end{eqnarray*}
where    
          \begin{eqnarray*}
          \omega = {\rm e}^{ {\rm i} \frac{2 \pi}{2j+1} } 
          \end{eqnarray*}
is a primitive $(2j+1)$-th root of unity and $a$ is a fixed parameter in the ring $\mathbb{Z}_{2j+1}$. The operator $v_a$ takes its origin in a polar decomposition of the two generators $E_{\pm} = J_{\pm}$ of the group SU$(2)$. For fixed $a$, the common eigenvectors of $J^2$ and $v_{a}$ provide an alternative basis to that given by the common eigenstates of $J^2$ and $J_z$. This can be made precise by the following result. 

\medskip 
\noindent $\blacktriangleright$~{\bf Proposition.} For fixed $j$ and $a$ (with $2j \in \mathbb{N}^*$ and 
$a \in \mathbb{Z}_{2j+1}$), the $2j+1$ common eigenvectors of $J^2$ and $v_{a}$ can be taken in the form
          \begin{eqnarray*}
|j \alpha ; a \rangle = \frac{1}{\sqrt{2j+1}} \sum_{m = -j}^{j} 
\omega^{\frac{1}{2}(j + m)(j - m + 1)a + (j + m)\alpha} | j , m \rangle 
          \end{eqnarray*} 
with $\alpha = 0, 1, \cdots, 2j$. The corresponding eigenvalues of $v_{a}$ are given by 
      \begin{eqnarray*}
v_{a} |j \alpha ; a \rangle = \omega^{ja - \alpha} |j \alpha ; a \rangle 
      \end{eqnarray*}
Then, the spectrum of $v_{a}$ is non-degenerate.~$\blacktriangleleft$ 

\medskip 
The inner product 
      \begin{eqnarray*}
 \langle j \alpha ;  a | j \beta ; a \rangle = \delta_{\alpha,\beta}, \quad \alpha, \beta = 0, 1, \cdots, 2j
      \label{jalphabetara}
      \end{eqnarray*}
shows that for fixed $j$ and $a$ 
      \begin{eqnarray*}
B_{a} = \{ |j \alpha ; a \rangle \ : \ \alpha = 0, 1, \cdots, 2j \}
      \label{Bra basis}
      \end{eqnarray*} 
is an orthonormal set which provides a nonstandard basis for the irreducible representation $(j)$ of SU$(2)$. For fixed $j$, there exists $2j+1$ orthonormal bases 
$B_{a}$ since $a$ can take $2j+1$ distinct values ($a = 0, 1, \cdots, 2j$). 

\subsubsection{Bases in quantum information} \label{Bases in quantum information}

We now go back to quantum information. By introducing 
$$
|a \alpha \rangle = |j \alpha ; a \rangle
$$
together with the change of notations (\ref{relationnjm}), the eigenvectors of $v_{a}$ can be written as  
      \begin{eqnarray*}
|a \alpha \rangle =
\frac{1}{\sqrt{d}} \sum_{n \in \mathbb{Z}_d}
\omega^{\frac{1}{2}(n+1)(d - n - 1) a - (n + 1) \alpha} \vert n \rangle
      \end{eqnarray*}
where $\omega = {\rm e}^{ {\rm i} \frac{2 \pi}{d} }$. The vector $|a \alpha \rangle$ satisfies the eigenvalue equation 
$$
v_a |a \alpha \rangle = \omega^{\frac{1}{2} (d-1) a - \alpha} |a \alpha \rangle
$$
For fixed $d$ and $a$, each eigenvector $|a \alpha \rangle$ is a linear combination of the qudits $| 0 \rangle, | 1 \rangle, \cdots, | d-1 \rangle$ and the basis
      \begin{eqnarray*}
B_{a} = \{ |a \alpha \rangle \ : \ \alpha = 0, 1, \cdots, d-1 \}
      \end{eqnarray*} 
is an alternative to the computational basis $B_d$. For fixed $d$, we therefore have $d+1$ remarkable bases of the $d$-dimensional space $\mathbb{C}^d$, namely, 
$B_d$ and $B_a$ for $a = 0, 1, \cdots, d-1$.

The operator $v_{a}$ can be represented by a $d$-dimensional unitary matrix ${V_{a}}$. The matrix ${V_{a}}$, built on the basis $B_d$ with the ordering 
$0, 1, \cdots, d-1$ for the lines and columns, reads 
      \begin{eqnarray*}
{{V_{a}}} =
\pmatrix{
0                      &    \omega^{a} &      0  & \cdots &          0    \cr
0                      &      0 & \omega^{2a}  & \cdots &          0    \cr
\vdots                 & \vdots & \vdots  & \cdots &     \vdots    \cr
0                      &      0 &      0  & \cdots & \omega^{(d-1)a}    \cr
1                      &      0 &      0  & \cdots &          0    \cr
}
      \end{eqnarray*}
The eigenvectors of ${V_{a}}$ are 
      \begin{eqnarray*}
\phi(a \alpha) = 
\frac{1}{\sqrt{d}} \sum_{n \in \mathbb{Z}_d}
\omega^{\frac{1}{2} (n + 1) (d - n - 1) a - (n + 1) \alpha} \phi_{n}
      \end{eqnarray*}
with $\alpha = 0, 1, \cdots, d-1$, where $\phi_{n}$ with $n = 0, 1, \cdots, d-1$ are the column vectors
      \begin{eqnarray*}	
\phi_0 = \pmatrix{
1      \cr
0      \cr
\vdots \cr
0
}, \quad 
\phi_1 =\pmatrix{
0      \cr
1      \cr
\vdots \cr
0
}, \quad 
\cdots, \quad 
\phi_{d-1}  = \pmatrix{
0      \cr
0      \cr
\vdots \cr
1
}
      \end{eqnarray*}
representing the qudits $| 0 \rangle, | 1 \rangle, \cdots, | d-1 \rangle$, respectively. The vectors $\phi(a \alpha)$  
satisfy the eigenvalue equation 
      \begin{eqnarray*}
{V_{a}} \phi(a \alpha) = \omega^{\frac{1}{2}(d-1)a - \alpha} \phi(a \alpha)
      \end{eqnarray*}
with the orthonormality relation 
$$
{\phi(a \alpha)}^{\dagger} \phi(a \beta) = \delta_{\alpha , \beta}
$$
for $\alpha, \beta = 0, 1, \cdots, d-1$. 

The matrix ${V_{a}}$ can be diagonalized by means of the $d$-dimensional matrix ${H_{a}}$ of elements
      \begin{eqnarray*}
\left( {H_{a}} \right)_{n \alpha} =
\frac{1}{\sqrt{d}} \omega^{\frac{1}{2} (n + 1) (d - n - 1) a - (n + 1)\alpha}                         
      \end{eqnarray*}
with the lines and columns of ${H_{a}}$ arranged from left to right and from 
top to bottom in the order $n, \alpha = 0, 1, \cdots, d-1$. Indeed, by introducing the $d \times d$ permutation matrix 
$$
P = \pmatrix{
1      & 0      & 0      & \cdots    & 0      & 0      \cr
0      & 0      & 0      & \cdots    & 0      & 1      \cr
0      & 0      & 0      & \cdots    & 1      & 0      \cr
\vdots & \vdots & \vdots & \cdots    & \vdots & \vdots \cr
0      & 0      & 1      & \cdots    & 0      & 0      \cr 
0      & 1      & 0      & \cdots    & 0      & 0      \cr   
}
$$
we can check that 
      \begin{eqnarray*}
(H_a P)^{\dagger} {V_a} (H_a P) =
\omega^{\frac{1}{2}(d-1)a} \pmatrix{
\omega^{0}           &      0     &       \cdots &          0   \cr
0                    & \omega^{1} &       \cdots &          0   \cr
\vdots               & \vdots     &       \cdots &     \vdots   \cr
0                    &      0     &       \cdots & \omega^{d-1} \cr
}
      \end{eqnarray*} 
from which we recover the eigenvalues of $V_a$. Note that the complex matrix $H_a$ is a unitary matrix for which each entry has a modulus equal to 
$\frac{1}{\sqrt{d}}$. Thus, $H_a$ is a generalized Hadamard matrix. This establishes a connection between MUBs and Hadamard matrices \cite{23Wocjan,35Aschbacher,36Bengtsson,45Brierley,44Kibler,Dita,52Goyeneche}. 

\subsubsection{MUBs for $d=p$ ($p$ prime)} \label{complete-set-of-MUBs}

Going back to the case where $d$ is arbitrary, we now examine an important property for the couple ($B_{a}, B_d$) and its 
generalization to couples ($B_{a}, B_{b}$) with $b \not= a$ ($a,b = 0, 1, \cdots, d-1$). For fixed $d$ and $a$, we verify that  
      \begin{eqnarray*}
\vert \langle n | a \alpha \rangle \vert = \frac{1}{\sqrt{d}}, \quad n, \alpha = 0, 1, \cdots, d-1 
      \end{eqnarray*}
which shows that $B_{a}$ and $B_d$ are two unbiased bases of the Hilbert space $\mathbb{C}^{d}$. 

Other examples of unbiased bases can be obtained for $d = 2$ and $3$. We easily check that the bases $B_{0}$ and $B_{1}$ for $d=2$ are unbiased. Similarly, the bases $B_{0}$, $B_{1}$ and $B_{2}$ for $d=3$ are mutually unbiased. Therefore, by taking into account the computational basis $B_d$, we end up with $d+1 = 3$ MUBs for $d=2$ and $d+1 = 4$ MUBs for $d=3$. This is in agreement with the general result according to which, in dimension $d$, the maximum number $d+1$ of MUBs is attained when $d$ is a prime number or a power of a prime number. The results for $d=2$ and $3$ can be generalized through the following proposition.  

\medskip 
\noindent $\blacktriangleright$~{\bf Proposition.} For $d=p$, $p$ a prime number, the bases $B_{0}, B_{1}, \cdots, B_{p}$ form a complete set of $p+1$ 
MUBs. The $p^2$ vectors $| a \alpha \rangle$, with $a, \alpha = 0, 1, \cdots, p-1$, of the bases $B_0, B_1, \cdots, B_{p-1}$ are given by a single formula, namely 
      \begin{eqnarray}
|a \alpha \rangle =  
\frac{1}{\sqrt{p}} \sum_{n \in \mathbb{F}_p}
\omega^{\frac{1}{2}(n+1)(p - n - 1) a - (n + 1) \alpha} \vert n \rangle, \quad \omega = {\rm e}^{ {\rm i} \frac{2 \pi}{p} }
			\label{masterformula}
			\end{eqnarray}
that gives the $p$ basis vectors for each basis $B_{a}$. In matrix form, $|a \alpha \rangle$ and $\vert n \rangle$ are replaced by $\phi(a \alpha)$ and $\phi_n$, respectively.~$\blacktriangleleft$ 

\medskip 
{\em Proof}. First, the computational basis $B_{p}$ is clearly unbiased to any of the $p$ bases $B_{0}, B_{1}, \cdots, B_{p-1}$. Second, let us consider
    \begin{eqnarray*}	                  
\langle a \alpha | b \beta \rangle = \frac{1}{p} 
\sum_{k = 0}^{p-1} {\rm e}^{{\rm i} \frac{\pi}{p} \{ (a-b)k^2 + [(b-a)p + 2(\beta - \alpha)]k \} }
     \end{eqnarray*}
for $b \not= a$. The inner product $\langle a \alpha | b \beta \rangle$ can be rewritten by making use of the generalized quadratic Gauss sum \cite{Berndt12}
     \begin{eqnarray*}
     S(u, v, w) = \sum_{k = 0}^{|w|-1} {\rm e}^{{\rm i} \frac{\pi}{w} (u k^2 + v k)} 
     \end{eqnarray*}
where $u$, $v$ and $w$ are integers such that $u$ and $w$ are co-prime, $uw$ is non-vanishing and $uw + v$ is even. This leads to 
     \begin{eqnarray*}
\langle a \alpha | b \beta \rangle = \frac{1}{p} S(u, v, w), \quad u = a - b, \quad v = -(a - b)p - 2(\alpha - \beta), \quad w = p
     \end{eqnarray*} 
It can be shown that $|S(u, v, w)| = \sqrt{p}$. Consequently  
     \begin{eqnarray*}
 | \langle a \alpha | b \beta \rangle | = \frac{1}{\sqrt{p}} 
     \end{eqnarray*}
for $b \not= a$ and $\alpha, \beta = 0, 1, \cdots, p-1$. This completes the proof.                \hfill $\square$ 

\medskip 
In many of the papers dealing with the construction of MUBs for $d = p$ a prime number or $d = p^m$ a power of a prime number, the explicit derivation of the bases requires the diagonalization of a set of matrices. The formula (\ref{masterformula}) arises from the diagonalization of a single matrix. It allows to derive in one step the $p(p + 1)$ vectors (or qupits, i.e.~qudits with $d = p$) of a complete set of $p + 1$ MUBs in $\mathbb{C}^p$ via a single formula easily encodable on a classical computer. 

Note that, for $d$ arbitrary, the inner product $\langle a \alpha | b \beta \rangle$ can be rewritten as
     \begin{eqnarray*}
\langle a \alpha | b \beta \rangle =      \left( {H_{a}}^{\dagger} H_{b}       \right)_{\alpha \beta} 
     \end{eqnarray*}
in terms of the generalized Hadamard matrices $H_a$ and $H_b$. In the case where $d = p$ is a prime number, we find that 
$$
\left| \left( {H_{a}}^{\dagger} H_{b}       \right)_{\alpha \beta} \right| = | \langle a \alpha | b \beta \rangle | = \frac{1}{\sqrt{p}} 
$$
Therefore, the product ${H_{a}}^{\dagger} H_{b}$ is another generalized Hadamard matrix \cite{44Kibler}.

Finally note that the passage, given by Eq.~(\ref{masterformula}), from the computational basis $B_p = \{ | n \rangle \ : \ n = 0, 1, \cdots, p-1 \}$ to the the basis $B_0 = \{ | 0 \alpha \rangle \ : \ \alpha  = 0, 1, \cdots, p-1 \}$ corresponds to a discrete Fourier transform. Similarly, the passage from the basis $B_p$ to the the basis $B_a = \{ | a \alpha \rangle \ : \ \alpha  = 0, 1, \cdots, p-1 \}$ with $a = 1, 2, \cdots, p-1$ corresponds to a {\em quadratic discrete Fourier transform}. 

{\bf {Example: {\rm $d = 2$}.}} In this case, relevant for a spin $j = \frac{1}{2}$ or for a qubit, we have $\omega = {\rm e}^{{\rm i} \pi}$ and $a, \alpha \in \mathbb{F}_2$. The matrices of the operators $v_{a}$ are 
     \begin{eqnarray}
V_{0} = 
\pmatrix{
  0     &1   \cr
  1     &0   \cr
} = \sigma_1, \quad 
V_{1} = 
\pmatrix{
  0     &-1  \cr
  1     &0   \cr
} = - i \sigma_2
     \nonumber 
     \end{eqnarray} 
The $d + 1 = 3$ MUBs $B_{0}$, $B_{1}$ and $B_{2}$ are the following:
\begin{eqnarray*}
B_{0}: & & 
| 0 0 \rangle =          \frac{| 0 \rangle +         | 1 \rangle}    {\sqrt{2}}      =  \frac{1}{\sqrt{2}} \pmatrix{
  1  \cr
  1  \cr
}, \quad
| 0 1 \rangle = -        \frac{| 0 \rangle -         | 1 \rangle}    {\sqrt{2}}      = - \frac{1}{\sqrt{2}} \pmatrix{
  1  \cr
 -1  \cr
} \\
B_{1}: & &
| 1 0 \rangle =  {\rm i} \frac{| 0 \rangle - {\rm i} | 1 \rangle}    {\sqrt{2}}      =   \frac{{\rm i}}{\sqrt{2}} \pmatrix{
  1  \cr
 -{\rm i}  \cr
}, \quad
| 1 1 \rangle = -{\rm i} \frac{| 0 \rangle + {\rm i} | 1 \rangle}    {\sqrt{2}}      = - \frac{{\rm i}}{\sqrt{2}} \pmatrix{
  1  \cr
  {\rm i}  \cr
} \\
B_{2} : & &
| 0 \rangle =
\pmatrix{
  1  \cr
  0  \cr
}, \quad
| 1 \rangle =
\pmatrix{
  0  \cr
  1  \cr
} 
\end{eqnarray*}
to be compared with Eq.~(\ref{B0B1B2}). 

{\bf {Example: {\rm $d = 3$}.}} This case corresponds to an angular momentum $j=1$ or to a qutrit. Here, we have 
$\omega = {\rm e}^{{\rm i} \frac{2 \pi}{3}}$ and $a, \alpha \in \mathbb{F}_3$. The matrices of the operators $v_{a}$ are 
\begin{eqnarray}
V_{0} = 
\pmatrix{
  0     &1   &0 \cr
  0     &0   &1 \cr
  1     &0   &0 \cr
}, \quad 
V_{1} =  
\pmatrix{
  0     &\omega   &0   \cr
  0     &0   &\omega^2 \cr
  1     &0   &0   \cr
}, \quad 
V_{2} = 
\pmatrix{
  0     &\omega^2   &0 \cr
  0     &0     &\omega \cr
  1     &0     &0 \cr
} \nonumber
\end{eqnarray}
The $d + 1 = 4$ MUBs $B_{0}$, $B_{1}$, $B_{2}$ and $B_{3}$ are the following:
\begin{eqnarray*}
B_{0} : & & | 0 0 \rangle = \frac{            | 0 \rangle +          | 1 \rangle + | 2 \rangle  } {\sqrt{3}}, \quad
            | 0 1 \rangle = \frac{   \omega^2 | 0 \rangle + \omega   | 1 \rangle + | 2 \rangle  } {\sqrt{3}}, \quad
            | 0 2 \rangle = \frac{   \omega   | 0 \rangle + \omega^2 | 1 \rangle + | 2 \rangle  } {\sqrt{3}} \\
B_{1} : & & | 1 0 \rangle = \frac{   \omega   | 0 \rangle + \omega   | 1 \rangle + | 2 \rangle  } {\sqrt{3}}, \quad
            | 1 1 \rangle = \frac{            | 0 \rangle + \omega^2 | 1 \rangle + | 2 \rangle  } {\sqrt{3}}, \quad
            | 1 2 \rangle = \frac{   \omega^2 | 0 \rangle +          | 1 \rangle + | 2 \rangle  } {\sqrt{3}} \\
B_{2} : & & | 2 0 \rangle = \frac{   \omega^2 | 0 \rangle + \omega^2 | 1 \rangle + | 2 \rangle  } {\sqrt{3}}, \quad
            | 2 1 \rangle = \frac{   \omega   | 0 \rangle +          | 1 \rangle + | 2 \rangle  } {\sqrt{3}}, \quad
            | 2 2 \rangle = \frac{            | 0 \rangle + \omega   | 1 \rangle + | 2 \rangle  } {\sqrt{3}} \\
B_{3} : & & | 0 \rangle,  \quad 
            | 1 \rangle,  \quad  
            | 2 \rangle   
\end{eqnarray*}
This can be transcribed in terms of column vectors as follows:  
\begin{eqnarray*}
B_{0} : & & 
| 0 0 \rangle =   \frac{1}{\sqrt{3}} \pmatrix{
  1  \cr
  1  \cr
  1  \cr
}, \
| 0 1 \rangle =   \frac{1}{\sqrt{3}} \pmatrix{
  \omega^2  \cr
  \omega    \cr
  1    \cr
}, \
| 0 2 \rangle =   \frac{1}{\sqrt{3}} \pmatrix{
  \omega   \cr
  \omega^2 \cr
  1   \cr
} \\
B_{1} : & &
| 1 0 \rangle =   \frac{1}{\sqrt{3}} \pmatrix{
  \omega  \cr
  \omega  \cr
  1  \cr
}, \
| 1 1 \rangle =   \frac{1}{\sqrt{3}} \pmatrix{
  1    \cr
  \omega^2  \cr
  1    \cr
}, \
| 1 2 \rangle =   \frac{1}{\sqrt{3}} \pmatrix{
  \omega^2    \cr
  1      \cr
  1      \cr
} \\
B_{2} : & &
| 2 0 \rangle =   \frac{1}{\sqrt{3}} \pmatrix{
  \omega^2  \cr
  \omega^2  \cr
  1    \cr
}, \
| 2 1 \rangle =   \frac{1}{\sqrt{3}} \pmatrix{
  \omega  \cr
  1  \cr
  1  \cr
}, \
| 2 2 \rangle =   \frac{1}{\sqrt{3}} \pmatrix{
  1  \cr
  \omega  \cr
  1  \cr
} \\ 
B_{3} : & &
| 0 \rangle = 
\pmatrix{
  1  \cr
  0  \cr
  0  \cr
}, \quad
| 1 \rangle = 
\pmatrix{
  0  \cr
  1  \cr
  0  \cr
}, \quad
| 2 \rangle = 
\pmatrix{
  0  \cr
  0  \cr
  1  \cr
} 
\end{eqnarray*}

To close this section, note that it is not necessary to treat separately the cases $p$ odd and $p$ even: the formula (\ref{masterformula}) for $| a \alpha \rangle$ is valid both for $p$ even prime ($p = 2$) and for $p$ odd prime. In the case where $p$ is odd, there exists a useful alternative formula to Eq.~(\ref{masterformula}) as shown in the next section.

\subsubsection{MUBs for $d = p$ ($p$ odd prime)} \label{alternative formula}

In the special case where $d = p$ is an odd prime number, the formula 
\begin{eqnarray}
| a \alpha \rangle' = \frac{1}{\sqrt{p}} \sum_{n \in \mathbb{F}_p} \omega^{(a n + \alpha) n} | n \rangle, \quad \omega 
                    = {\rm e}^{{\rm i} \frac{2 \pi}{p}} 
										\label{alternativef}
\end{eqnarray}
provides an alternative to the formula (\ref{masterformula}). Indeed, it can be shown that 
$$
{B_a}' =  \{ | a \alpha \rangle' \ : \ \alpha = 0, 1, \cdots, p-1 \}
$$
where $a$ can take any of the values $0, 1, \cdots, p-1$ constitutes an orthonormal basis of $\mathbb{C}^d$ and that the $p$ bases ${B_a}'$ ($a = 0, 1, \cdots, p-1$) form, with the computational basis $B_p$, a complete set of $p+1$ MUBs. The proof, based on the properties of Gauss sums, is analogous to that given in Section \ref{complete-set-of-MUBs}. 

It is to be emphasized that for $p$ even prime ($p= 2$) the bases ${B_0}'$, ${B_1}'$ and $B_2$ do not form a complete set of MUBs while the proposition given in Section \ref{complete-set-of-MUBs} is valid for $p$ odd prime and equally well for $p$ even prime. The interest of Eq.~(\ref{alternativef}) is that it can be easily extended in the case where $\mathbb{F}_p$ is replaced by the Galois field $\mathbb{GF}(p^m)$ with $m > 1$. 

\subsubsection{MUBs for $d$ power of a prime} \label{MUBs by tensor product}

We may ask what becomes the proposition in Section \ref{complete-set-of-MUBs} when the prime number $p$ is replaced by an arbitrary (not prime) number $d$. In this case, the formula (\ref{masterformula}), with $p$ replaced by $d$, does not provide a complete set of $d + 1$ MUBs. However, it is easy to verify that the bases $B_{0}$, $B_{1}$ and 
$B_d$ are three MUBs in $\mathbb{C}^d$, in agreement with the well-known result according to which the number of MUBs in $\mathbb{C}^d$, with $d$ arbitrary, is greater than or equal to 3. 

The formula (\ref{masterformula}) for $\mathbb{C}^p$ can be used for deriving a complete set of $p^m + 1$ MUBs in $\mathbb{C}^{p^m}$ ($p$ prime and $m \geq 2$) by tensor products of order $m$ of vectors in $\mathbb{C}^p$. The general case is very much involved. Hence, we shall limit ourselves to the case $d = 2^2$.  

The case $d = 4$ corresponds to the spin angular momentum $j = \frac{3}{2}$. The four bases $B_a$ for $a = 0, 1, 2, 3$ consisting of the vectors $| a \alpha \rangle$ calculated for $d = 4$ from Section \ref{Bases in quantum information} and the computational basis $B_4$ do not constitute a complete set of $d+1 = 5$ MUBs. Nevertheless, it is possible to find $d +1 = 5$ MUBs because $d = 2^2$ is the power of a prime number. Indeed, another way to deal with the search for MUBs in 
$\mathbb{C}^4$ is to consider two systems of qubits associated with the spin angular momenta $j_1 = \frac{1}{2} \Leftrightarrow d_1 = p = 2$ and 
$j_2 = \frac{1}{2} \Leftrightarrow d_2 = p = 2$. Then, bases of $\mathbb{C}^4$ can be constructed from tensor products 
$|a \alpha \rangle \otimes |b \beta \rangle$ which are eigenvectors of the operator $v_{a} \otimes v_{b}$, where $v_a$ corresponds to the first system of qubits and 
$v_b$ to the second one. Obviously, the set 
   	\begin{eqnarray}
B_{ab} = \{ |a \alpha \rangle \otimes |b \beta \rangle \ : \ \alpha, \beta = 0, 1 \}
    \nonumber
   	\end{eqnarray} 
is an orthonormal basis of $\mathbb{C}^4$. Four of the five MUBs for $d = 2^2 = 4$ can be constructed from the various bases $B_{ab}$. It is evident that $B_{00}$ and $B_{11}$ are two unbiased bases since the modulus of the inner product of $|1 \alpha' \rangle \otimes |1 \beta' \rangle$ by 
$|0 \alpha  \rangle \otimes |0 \beta  \rangle$ is 
   	\begin{eqnarray}
| \langle 0 \alpha | 1 \alpha' \rangle \langle 0 \beta | 1 \beta' \rangle | = \frac{1}{\sqrt{4}} = \frac{1}{\sqrt{d}}
    \nonumber
   	\end{eqnarray}    	
A similar result holds for the two bases $B_{01}$ and $B_{10}$. However, the four bases $B_{00}$, $B_{11}$, $B_{01}$ and $B_{10}$ are not mutually unbiased. A possible way to overcome this no-go result is to keep the bases $B_{00}$ and $B_{11}$ intact and to re-organize the vectors inside the bases $B_{01}$ and $B_{10}$ in order to obtain four MUBs. We are thus left with the four bases 	
   	\begin{eqnarray}
W_{00} \equiv B_{00}, \quad W_{11} \equiv B_{11}, \quad W_{01}, \quad W_{10}
    \nonumber
   	\end{eqnarray}   
which together with the computational basis $B_4$ give five MUBs. In detail, we have 
   	\begin{eqnarray}
W_{00} & = & \{         |0 \alpha \rangle \otimes |0 \beta \rangle \ : \ \alpha, \beta = 0, 1 \} \nonumber \\
W_{11} & = & \{         |1 \alpha \rangle \otimes |1 \beta \rangle \ : \ \alpha, \beta = 0, 1 \} \nonumber \\
W_{01} & = & \{ \lambda |0 \alpha \rangle \otimes |1 \beta \rangle + 
                     \mu |0 \alpha \oplus 1 \rangle \otimes |1 \beta \oplus 1 \rangle \ : \ \alpha, \beta = 0, 1 \} \nonumber \\
W_{10} & = & \{ \lambda |1 \alpha \rangle \otimes |0 \beta \rangle + 
                     \mu |1 \alpha \oplus 1 \rangle \otimes |0 \beta \oplus 1 \rangle \ : \ \alpha, \beta = 0, 1 \} \nonumber 
   	\end{eqnarray} 
where the addition $\oplus$ should be understood modulo 4; furthermore 
   	\begin{eqnarray}
\lambda = \frac{1-{\rm i}}{2}, \quad \mu = \frac{1+{\rm i}}{2}
    \nonumber
   	\end{eqnarray} 
and the vectors of type $|a \alpha \rangle$ are given by the formula (\ref{masterformula}). As a r\'esum\'e, only two formulas are necessary for obtaining the $d^2 = 16$ vectors $| a b ; \alpha \beta \rangle$ for the bases $W_{ab}$, namely 
   	  \begin{eqnarray*}
W_{00}, W_{11} &:& | a a ; \alpha \beta \rangle = |a \alpha \rangle \otimes |a \beta \rangle \\
W_{01}, W_{10} &:& | a a \oplus 1 ; \alpha \beta \rangle = \lambda |a \alpha          \rangle \otimes |a \oplus 1 \beta \rangle + 
                                                                 \mu |a \alpha \oplus 1 \rangle \otimes |a \oplus 1 \beta \oplus 1 \rangle 
   	  \end{eqnarray*} 
for all $a, \alpha, \beta$ in $\mathbb{F}_2$. A simple development of $W_{00}$, $W_{11}$, $W_{01}$ and $W_{10}$ gives the following expressions. 

\medskip
\emph{The $W_{00}$ basis}:
\begin{eqnarray*}
| 0 0 ; 0 0 \rangle &=& \frac{1}{2} 
(| 0 \rangle \otimes | 0 \rangle + | 0 \rangle \otimes | 1 \rangle + | 1 \rangle \otimes | 0 \rangle + | 1 \rangle \otimes | 1 \rangle) \nonumber \\
| 0 0 ; 0 1 \rangle &=& \frac{1}{2} 
(| 0 \rangle \otimes | 0 \rangle - | 0 \rangle \otimes | 1 \rangle + | 1 \rangle \otimes | 0 \rangle - | 1 \rangle \otimes | 1 \rangle) \nonumber \\
| 0 0 ; 1 0 \rangle &=& \frac{1}{2} 
(| 0 \rangle \otimes | 0 \rangle + | 0 \rangle \otimes | 1 \rangle - | 1 \rangle \otimes | 0 \rangle - | 1 \rangle \otimes | 1 \rangle) \nonumber \\
| 0 0 ; 1 1 \rangle &=& \frac{1}{2} 
(| 0 \rangle \otimes | 0 \rangle - | 0 \rangle \otimes | 1 \rangle - | 1 \rangle \otimes | 0 \rangle + | 1 \rangle \otimes | 1 \rangle) \nonumber      
\end{eqnarray*}
or in column vectors
	\begin{eqnarray}
\frac{1}{2} \pmatrix{
1 \cr
1 \cr
1 \cr
1 \cr
}, \quad
\frac{1}{2} \pmatrix{
1 \cr
-1 \cr
1 \cr
-1 \cr
}, \quad
\frac{1}{2} \pmatrix{
1 \cr
1 \cr
-1 \cr
-1 \cr
}, \quad
\frac{1}{2} \pmatrix{
1 \cr
-1 \cr
-1 \cr
1 \cr
}
	\nonumber 
	\end{eqnarray}
	
\emph{The $W_{11}$ basis}:
   \begin{eqnarray}
| 1 1 ; 0 0 \rangle &=& \frac{1}{2} 
(| 0 \rangle \otimes | 0 \rangle + {\rm i} | 0 \rangle \otimes | 1 \rangle + {\rm i} | 1 \rangle \otimes | 0 \rangle - | 1 \rangle \otimes | 1 \rangle) \nonumber  \\
| 1 1 ; 0 1 \rangle &=& \frac{1}{2} 
(| 0 \rangle \otimes | 0 \rangle - {\rm i} | 0 \rangle \otimes | 1 \rangle + {\rm i} | 1 \rangle \otimes | 0 \rangle + | 1 \rangle \otimes | 1 \rangle) \nonumber  \\
| 1 1 ; 1 0 \rangle &=& \frac{1}{2} 
(| 0 \rangle \otimes | 0 \rangle + {\rm i} | 0 \rangle \otimes | 1 \rangle - {\rm i} | 1 \rangle \otimes | 0 \rangle + | 1 \rangle \otimes | 1 \rangle) \nonumber  \\
| 1 1 ; 1 1 \rangle &=& \frac{1}{2} 
(| 0 \rangle \otimes | 0 \rangle - {\rm i} | 0 \rangle \otimes | 1 \rangle - {\rm i} | 1 \rangle \otimes | 0 \rangle - | 1 \rangle \otimes | 1 \rangle) \nonumber  
   \end{eqnarray}
or in column vectors
   \begin{eqnarray}
\frac{1}{2} \pmatrix{
1 \cr
{\rm i} \cr
{\rm i} \cr
-1 \cr
}, \quad
\frac{1}{2} \pmatrix{
1 \cr
-{\rm i} \cr
{\rm i} \cr
1 \cr
}, \quad
\frac{1}{2} \pmatrix{
1 \cr
{\rm i} \cr
-{\rm i} \cr
1 \cr
}, \quad
\frac{1}{2} \pmatrix{
1 \cr
-{\rm i} \cr
-{\rm i} \cr
-1 \cr
}
   \nonumber 
   \end{eqnarray}
	
\emph{The $W_{01}$ basis}:
   \begin{eqnarray}
| 0 1 ; 0 0 \rangle &=& \frac{1}{2} 
(| 0 \rangle \otimes | 0 \rangle + | 0 \rangle \otimes | 1 \rangle - {\rm i} | 1 \rangle \otimes | 0 \rangle + {\rm i} | 1 \rangle \otimes | 1 \rangle) \nonumber  \\
| 0 1 ; 1 1 \rangle &=& \frac{1}{2} 
(| 0 \rangle \otimes | 0 \rangle - | 0 \rangle \otimes | 1 \rangle + {\rm i} | 1 \rangle \otimes | 0 \rangle + {\rm i} | 1 \rangle \otimes | 1 \rangle) \nonumber  \\
| 0 1 ; 0 1 \rangle &=& \frac{1}{2} 
(| 0 \rangle \otimes | 0 \rangle - | 0 \rangle \otimes | 1 \rangle - {\rm i} | 1 \rangle \otimes | 0 \rangle - {\rm i} | 1 \rangle \otimes | 1 \rangle) \nonumber  \\
| 0 1 ; 1 0 \rangle &=& \frac{1}{2} 
(| 0 \rangle \otimes | 0 \rangle + | 0 \rangle \otimes | 1 \rangle + {\rm i} | 1 \rangle \otimes | 0 \rangle - {\rm i} | 1 \rangle \otimes | 1 \rangle) \nonumber      
\end{eqnarray}
or in column vectors
	\begin{eqnarray}
\frac{1}{2} \pmatrix{
1 \cr
1 \cr
-{\rm i} \cr
{\rm i} \cr
}, \quad
\frac{1}{2} \pmatrix{
1 \cr
-1 \cr
{\rm i} \cr
{\rm i} \cr
}, \quad
\frac{1}{2} \pmatrix{
1 \cr
-1 \cr
-{\rm i} \cr
-{\rm i} \cr
}, \quad
\frac{1}{2} \pmatrix{
1 \cr
1 \cr
{\rm i} \cr
-{\rm i} \cr
}
   \nonumber 
   \end{eqnarray}
	
\emph{The $W_{10}$ basis}:
   \begin{eqnarray}
| 1 0 ; 0 0 \rangle &=& \frac{1}{2} 
(| 0 \rangle \otimes | 0 \rangle - {\rm i} | 0 \rangle \otimes | 1 \rangle + | 1 \rangle \otimes | 0 \rangle + {\rm i} | 1 \rangle \otimes | 1 \rangle) \nonumber \\
| 1 0 ; 1 1 \rangle &=& \frac{1}{2} 
(| 0 \rangle \otimes | 0 \rangle + {\rm i} | 0 \rangle \otimes | 1 \rangle - | 1 \rangle \otimes | 0 \rangle + {\rm i} | 1 \rangle \otimes | 1 \rangle) \nonumber \\
| 1 0 ; 0 1 \rangle &=& \frac{1}{2} 
(| 0 \rangle \otimes | 0 \rangle + {\rm i} | 0 \rangle \otimes | 1 \rangle + | 1 \rangle \otimes | 0 \rangle - {\rm i} | 1 \rangle \otimes | 1 \rangle) \nonumber \\
| 1 0 ; 1 0 \rangle &=& \frac{1}{2} 
(| 0 \rangle \otimes | 0 \rangle - {\rm i} | 0 \rangle \otimes | 1 \rangle - | 1 \rangle \otimes | 0 \rangle - {\rm i} | 1 \rangle \otimes | 1 \rangle) \nonumber 
   \end{eqnarray}
or in column vectors
   \begin{eqnarray}
\frac{1}{2} \pmatrix{
1 \cr
-{\rm i} \cr
1 \cr
{\rm i} \cr
}, \quad
\frac{1}{2} \pmatrix{
1 \cr
{\rm i} \cr
-1 \cr
{\rm i} \cr
}, \quad
\frac{1}{2} \pmatrix{
1 \cr
{\rm i} \cr
1 \cr
-{\rm i} \cr
}, \quad
\frac{1}{2} \pmatrix{
1 \cr
-{\rm i} \cr
-1 \cr
-{\rm i} \cr
}
   \nonumber 
   \end{eqnarray}

\emph{The computational basis}:
	\begin{eqnarray}
| 0 \rangle \otimes | 0 \rangle, \quad 
| 0 \rangle \otimes | 1 \rangle, \quad 
| 1 \rangle \otimes | 0 \rangle, \quad 
| 1 \rangle \otimes | 1 \rangle
	\nonumber 
	\end{eqnarray} 	
or in column vectors	
	\begin{eqnarray*}
\pmatrix{
1 \cr
0 \cr
0 \cr
0 \cr
}, \quad
\pmatrix{
0 \cr
1 \cr
0 \cr
0 \cr
}, \quad
\pmatrix{
0 \cr
0 \cr
1 \cr
0 \cr
}, \quad
\pmatrix{
0 \cr
0 \cr
0 \cr
1 \cr
}
   \end{eqnarray*}

It is to be noted that the vectors of the bases $W_{00}$ and $W_{11}$ are not entangled (i.e.~each vector is the tensor product of two vectors) while the vectors of the bases $W_{01}$ and $W_{10}$ are entangled (i.e.~each vector is not the tensor product of two vectors). In fact, all the state vectors for $W_{01}$ and $W_{10}$ are maximally entangled (the entanglement entropy is maximum for $W_{01}$ and $W_{10}$ and vanishes for $W_{00}$ and 
$W_{11}$). 

Generalization of the formulas given above for two systems of qubits can be obtained in more complicated situations (two systems of qupits, three systems of qubits, etc.). The generalization of the bases $W_{00}$ and $W_{11}$ is immediate. The generalization of $W_{01}$ and $W_{10}$ can be achieved by taking linear combinations of vectors such that each linear combination is made of vectors corresponding to the same eigenvalue of the relevant tensor product of operators of type $v_{a}$.

\subsection{Weyl pairs} \label{Weyl}

\subsubsection{Shift and phase operators} \label{Weylpair}

Let us go back to the case $d$ arbitrary. The matrix ${V_{a}}$ can be decomposed as 
     \begin{eqnarray*}
{V_{a}} = {X} {Z}^a, \quad a = 0, 1, \cdots, d - 1
     \end{eqnarray*}
where
        \begin{eqnarray*}
{X} = 
\pmatrix{
0                    &      1 &      0  & \cdots &       0 \cr
0                    &      0 &      1  & \cdots &       0 \cr
\vdots               & \vdots & \vdots  & \cdots &  \vdots \cr
0                    &      0 &      0  & \cdots &       1 \cr
1                    &      0 &      0  & \cdots &       0 \cr
}, \quad 
{Z} = 
\pmatrix{
1                    &      0 &      0    & \cdots &       0       \cr
0                    &      \omega &      0    & \cdots &       0       \cr
0                    &      0 &      \omega^2  & \cdots &       0       \cr
\vdots               & \vdots & \vdots    & \cdots &  \vdots       \cr
0                    &      0 &      0    & \cdots &       \omega^{d-1} \cr
}, \quad \omega = {\rm e}^{{\rm i} \frac{2 \pi}{d}}
        \end{eqnarray*}
The matrices $X$ and $Z$ satisfy 
\begin{eqnarray*}
Z \phi_n = \omega^n \phi_n,     \ n = 0, 1, \cdots, d-1, \quad  
X \phi_n = \phi_{n-1 \> {\rm mod} \> d} = \cases{\phi_{d-1},   \ n = 0   \cr \cr 
                                                 \phi_{n - 1}, \ n = 1, 2, \cdots, d-1
									                              }
\end{eqnarray*}
The linear operators corresponding to the matrices ${X}$ and ${Z}$ are known in quantum information as flip or shift and clock or phase 
operators, respectively. The unitary matrices ${X}$ and ${Z}$ $\omega$-commute in the sense that 
     \begin{eqnarray*}
{X} {Z} - \omega {Z} {X} = {O_d}
     \end{eqnarray*}
In addition, they satisfy
     \begin{eqnarray*}
{X}^d = {Z}^d = {I_d} 
     \end{eqnarray*}
where ${I_d}$ and ${O_d}$ are the $d$-dimensional unity and zero matrices, respectively. The last two equations show that ${X}$ and ${Z}$ constitute a so-called Weyl pair \cite{Weyl}. 

Note that the Weyl pair ($X , Z$) can be deduced from the master matrix $V_a$ via
     \begin{eqnarray*}
{X} = {V_{0}}, \quad 
{Z} = {V_{0}}^{\dagger} {V_{1}}
     \end{eqnarray*} 
which shows a further interest of the matrix $V_{a}$. Indeed, the matrix $V_{a}$ condensates all that can be done with the matrices $X$ and $Z$. This has been seen in Section \ref{complete-set-of-MUBs} with the derivation of a single formula for the determination from $V_{a}$ of a complete set of $p+1$ MUBs when $d = p$ is  prime whereas many other determinations of such a complete set needs repeated use of the matrices $X$ and $Z$.

A connection between $X$ and $Z$ can be deduced from the expression of $(H_a P)^{\dagger} {V_a} (H_a P)$ given in \ref{Bases in quantum information}. By taking 
$a = 0$, we obtain 
$$
(H_0 P)^{\dagger} {X} (H_0 P) = Z \ \Leftrightarrow \ X = (H_0 P) Z (H_0 P)^{\dagger}
$$
where $H_0$ is the matrix of a discrete Fourier transform that allows to pass from the vectors $\phi_n$ ($n = 0, 1, \cdots, d-1$) to the vector 
$\phi(0, \alpha)$ according to 
$$
\phi(0, \alpha) = \sum_{n \in \mathbb{Z}_d} \left( H_0 \right)_{n \alpha} \phi_n 
                = (-1)^{\alpha} \frac{1}{\sqrt{d}} \sum_{n \in \mathbb{Z}_d} {\rm e}^{-{\rm i} \frac{2 \pi}{d} n \alpha} \phi_n
$$
cf.~the expression of $\phi(a, \alpha)$ in \ref{Bases in quantum information}. 

\subsubsection{Generalized Pauli matrices}

For $d$ arbitrary, let us define the matrices 
     \begin{eqnarray}
U_{ab} = X^a Z^b, \quad a, b \in \mathbb{Z}_d 
     \nonumber  
     \end{eqnarray}
The matrices $U_{ab}$ belong to the unitary group U$(d)$. The $d^2$ matrices $U_{ab}$ are called {\em generalized Pauli matrices} in dimension $d$. They satisfy the trace relation 
          \begin{eqnarray}
 {\rm tr} \left( {U_{ab}}^{\dagger} U_{a'b'} \right) = d \> \delta_{a,a'} \> \delta_{b,b'}
          \nonumber    
          \end{eqnarray}
Thus, the set $\{ U_{ab} \ : \ a, b \in \mathbb{Z}_d \}$ of unitary matrices is an orthogonal set with respect to the Hilbert-Schmidt inner product. Consequently, the 
$d^2$ pairwise orthogonal matrices $U_{ab}$ can be used as a basis of $\mathbb{C}^{d \times d}$.

\medskip
{\bf Example 1.} The case $d = 2 \Leftrightarrow j = \frac{1}{2}$ ($\Rightarrow \omega = {\rm e}^{{\rm i} \pi}$ and $a, b = 0,1$) corresponds to the two-dimensional ordinary Pauli matrices of quantum mechanics. The matrices $X^a Z^b$ are 
$$
I_2 = X^0 Z^0 = 
\pmatrix{
  1     &0   \cr
  0     &1   \cr
}, \ 
X = X^1 Z^0 = 
\pmatrix{
  0     &1   \cr
  1     &0   \cr
}, \ 
 Z = X^0 Z^1 = 
\pmatrix{
  1     &0   \cr
  0     &-1  \cr
}, \
Y = X^1 Z^1 = 
\pmatrix{
  0     &-1  \cr
  1     &0   \cr
} \nonumber
$$ 		
so that the matrices $X$ and $Z$ generate the ordinary Pauli matrices. Indeed, we have  
     \begin{eqnarray*}
I_2 =       \sigma_0,          \quad  
X = V_{0} = \sigma_1,          \quad   
Y = XZ = V_{1} = - i \sigma_2, \quad 
Z = \sigma_3   
     \end{eqnarray*}
in terms of the usual (Hermitian and unitary) Pauli matrices.

\medskip
{\bf Example 2.} The case $d = 3 \Leftrightarrow j = 1$ ($\Rightarrow \omega = {\rm e}^{{\rm i} \frac{2 \pi}{3}}$ and $a, b = 0,1,2$) yields nine three-dimensional matrices. More precisely, the matrices $X$ and $Z$ generate $I_3 = X^0 Z^0$ and 
     \begin{eqnarray}
X = V_{0}, \quad 
X^2, \quad 
Z, \quad 
Z^2, \quad 
XZ = V_{1}, \quad 
X^2Z^2, \quad 
XZ^2 = V_{2}, \quad
X^2Z
     \nonumber    
     \end{eqnarray} 
In the detail, the matrices $X^a Z^b$ are 
     \begin{eqnarray}
&& X^0 Z^0 =  
\pmatrix{
   1     &0     &0   \cr
   0     &1     &0   \cr
   0     &0     &1   \cr
 }, \ 
X^0 Z^1 = 
\pmatrix{
  1     &0     &0     \cr
  0     &\omega     &0     \cr
  0     &0     &\omega^2   \cr
}, \ 
X^0 Z^2 =  
\pmatrix{
  1     &0       &0   \cr
  0     &\omega^2     &0   \cr
  0     &0       &\omega   \cr
} \nonumber \\
&& X^1 Z^0 = 
\pmatrix{
  0     &1     &0   \cr
  0     &0     &1   \cr
  1     &0     &0   \cr
}, \ 
X^1 Z^1 = 
\pmatrix{
  0     &\omega     &0     \cr
  0     &0     &\omega^2   \cr
  1     &0     &0     \cr
}, \ 
X^1 Z^2 = 
\pmatrix{
  0     &\omega^2     &0     \cr
  0     &0       &\omega     \cr
  1     &0       &0     \cr
} \nonumber \\ 
&& X^2 Z^0 =  
\pmatrix{
  0     &0     &1   \cr
  1     &0     &0   \cr
  0     &1     &0   \cr
}, \
X^2 Z^1 = 
\pmatrix{
  0     &0     &\omega^2     \cr
  1     &0     &0       \cr
  0     &\omega     &0       \cr
}, \ 
X^2 Z^2 = 
\pmatrix{
  0     &0       &\omega     \cr
  1     &0       &0     \cr
  0     &\omega^2     &0     \cr
} \nonumber  
     \end{eqnarray}
They constitute a natural extension in dimension $d = 3$ of the usual Pauli matrices. 

\subsubsection{Weyl pair and groups}

For arbitrary $d$, the Weyl pair ($X=V_{0}$, $Z=V_{0}^{\dagger} V_{1}$) is a basic ingredient for generating the Pauli group $P_d$ in $d$ dimensions and the Lie algebra of the linear group GL$(d , \mathbb{C})$ in $d$ dimensions, groups of central interest in group theory, quantum mechanics and quantum information.     

\medskip
{\bf The Pauli group.} For arbitrary $d$, let us define the matrices 
                       \begin{eqnarray}
V_{abc} = \omega^a U_{bc} = \omega^a X^b Z^c, \quad a, b, c \in \mathbb{Z}_d, \quad \omega = {\rm e}^{{\rm i} \frac{2 \pi}{d}}
                       \nonumber
                       \end{eqnarray} 
The matrices $V_{abc}$ are unitary and satisfy 
                       \begin{eqnarray}
{\rm tr} \left( {V_{abc}}^{\dagger} V_{a'b'c'} \right) = \omega^{a' - a} \> d \> \delta_{b,b'} \> \delta_{c,c'}
                       \nonumber
                       \end{eqnarray} 
In addition, we have the following result.

\medskip 
\noindent $\blacktriangleright$~{\bf Proposition.} The set $\{ V_{abc} \ : \ a, b, c \in \mathbb{Z}_d \}$ is a finite group of order $d^3$, denoted $P_d$, for the internal law (matrix multiplication)
\begin{eqnarray}
V_{a b c} V_{a' b' c'} = V_{a'' b'' c''}, \quad 
                          a'' = a+a'-cb', \quad 
                              b'' = b+b', \quad 
                              c'' = c+c' 
\nonumber
\end{eqnarray} 
It is a non-commutative (for $d \geq 2$) nilpotent group with nilpotency class equal to 3.~$\blacktriangleleft$ 

\medskip
The group $P_d$ is called the Pauli group in dimension $d$. It is of considerable importance in quantum information, especially for quantum computation and for quantum error-correcting codes. The group $P_d$ is a sub-group of the unitary group U$(d)$. The normalizer of $P_d$ in U$(d)$ is called Clifford group (denoted as ${\cal C}_d$) in $d$ dimensions. More precisely, ${\cal C}_d$ is the set $\{ U \in {\rm U}(d) \ : \ U P_d U^{\dagger} = P_d \}$ endowed with matrix multiplication. The Pauli group $P_d$ as well as any other invariant sub-group of ${\cal C}_d$ are useful for quantum error-correcting codes in the case of $N$-qubit systems corresponding to $d = 2^N$. 

Moreover, the Pauli group is connected to the Heisenberg-Weyl group. In fact, the group $P_d$ corresponds to a discretization of the Heisenberg-Weyl group $HW(\mathbb{R})$. From an abstract point of view, the group $HW(\mathbb{R})$ is the set $S = \{ (x, y, z) \ : \ x, y, z \in \mathbb{R} \}$ equipped with the internal law $S \times S \to S$ defined via 
          \begin{eqnarray}
(x, y, z) (x', y', z') = (x + x' - zy', y + y', z + z')
          \nonumber    
          \end{eqnarray} 
This group is a non-commutative Lie group of order 3. It is non-compact and nilpotent with a nilpotency class equal to 3. The passage from $HW(\mathbb{R})$ to $P_d$ amounts to replace the infinite field $\mathbb{R}$ by the finite ring $\mathbb{Z}_d$ so that $HW(\mathbb{R})$ gives $HW(\mathbb{Z}_d) \equiv P_d$. 

The three generators of $HW(\mathbb{R})$ are
          \begin{eqnarray}
{H} = \frac{1}{{\rm i}}        \frac{\partial}{\partial x}, \quad 
{Q} = \frac{1}{{\rm i}}        \frac{\partial}{\partial y}, \quad 
{P} = \frac{1}{{\rm i}} \left( \frac{\partial}{\partial z} - y \frac{\partial}{\partial x} \right)
          \nonumber    
          \end{eqnarray} 
They satisfy the commutation relations  
          \begin{eqnarray}
[Q , P]_- = {\rm i} H, \quad 
[P , H]_- = 0,   \quad 
[H , Q]_- = 0
          \nonumber    
          \end{eqnarray} 
Therefore, the Lie algebra $hw(\mathbb{R})$ of $HW(\mathbb{R})$ is a three-dimensional nilpotent Lie algebra with nilpotency class equal to 3. The commutation relations of $Q$, $P$ and $H$ are reminiscent of the Heisenberg commutation relations. As a matter of fact, the Heisenberg commutation relations correspond to an infinite-dimensional irreducible representation by Hermitian matrices of $hw(\mathbb{R})$. The Lie algebra 
$hw(\mathbb{R})$ also admits finite-dimensional irreducible representations at the price to abandon the Hermitian character of the representation matrices. 

\medskip
{\bf The linear group.} The Weyl pair consisting of the generalized Pauli matrices $X$ and $Z$ in $d$ dimensions can be used for constructing a basis of the Lie algebra of U$(d)$. More precisely, we have the two following propositions. 

\medskip 
\noindent $\blacktriangleright$~{\bf Proposition.} For arbitrary $d$, the set $\{ X^a Z^b \ : \ a , b \in \mathbb{Z}_d \}$ forms a basis for the Lie algebra 
gl$(d, \mathbb{C})$ of the linear group GL$(d, \mathbb{C})$ or for the Lie algebra u$(d)$ of the unitary group U$(d)$. The Lie brackets of 
gl$(d, \mathbb{C})$ in such a basis are 
     \begin{eqnarray}
[ X^a Z^b , X^{e} Z^{f} ]_- = \sum_{i \in \mathbb{Z}_d} \sum_{j \in \mathbb{Z}_d} (ab,ef;ij) X^{i} Z^{j}
     \nonumber    
     \end{eqnarray}
with the structure constants  
      \begin{eqnarray}
(ab,ef;ij) = \delta_{i, a+e} 
             \delta_{j, b+f}
		 \left( \omega^{- be} - \omega^{- af} \right) 
     \nonumber    
     \end{eqnarray}
where $a, b, e, f, i, j \in \mathbb{Z}_d$.~$\blacktriangleleft$ 

\medskip 
Note that the commutator $[U_{ab}, U_{ef}]_- = U_{ab} U_{ef} - U_{ef} U_{ab}$ and the anti-commutator 
                         $[U_{ab}, U_{ef}]_+ = U_{ab} U_{ef} + U_{ef} U_{ab}$ of $U_{ab}$ and $U_{ef}$ are given by
                  \begin{eqnarray}
[U_{ab}, U_{ef}]_{\pm}= \left( \omega^{-be} \pm \omega^{-af} \right) U_{ij}, \quad i = a + e, \quad j = b + f
                  \nonumber
                  \end{eqnarray}
Consequently, $[U_{ab}, U_{ef}]_- = 0$ if and only if $af - be = 0$ (mod $d$) and $[U_{ab}, U_{ef}]_+ = 0$ if and only if $af - be = \frac{1}{2} d$ (mod $d$). Therefore, all anti-commutators $[U_{ab}, U_{ef}]_+$ are different from 0 if $d$ is an odd integer.

\medskip 
\noindent $\blacktriangleright$~{\bf Proposition.} For $d=p$, with $p$ a prime number, the simple Lie algebra sl$(p, \mathbb{C})$ of the special linear group SL$(p, \mathbb{C})$ or its compact real form su$(d)$ of the special unitary group SU$(d)$ can be decomposed into a sum of 
$p+1$ Abelian subalgebras of dimension $p-1$ 
                  \begin{eqnarray}
{\rm sl}(p, \mathbb{C}) = 
{\cal V}_0     \oplus 
{\cal V}_1     \oplus 
\cdots     \oplus      
{\cal V}_{p}     
                  \nonumber
                  \end{eqnarray}
where each of the $p+1$ subalgebras ${\cal V}_0, {\cal V}_1, \cdots, {\cal V}_p$ is a Cartan subalgebra generated by a set of $p - 1$ commuting matrices.~$\blacktriangleleft$ 

\medskip
A similar result holds for $d = p^m$, a power of a prime number \cite{Kostrikin,08Patera,Kostrikinlivre,37Boykin,44Kibler}. 

The decomposition of sl$(p, \mathbb{C})$, called orthogonal decomposition of sl$(p, \mathbb{C})$, is trivial for $p = 2$. In fact, for $p= 2$ we have the following decomposition 
$$
{\rm su}(2) = \sigma_1 \oplus \sigma_2 \oplus \sigma_3 
$$
in terms of vector space sum.

\subsubsection{MUBs and the special linear group}

According to the orthogonal decomposition proposition, in the case where $d = p$ is a prime number (even or odd), the 
set $\{ {X}^a{Z}^b \ : \ a,b \in \mathbb{Z}_p \} \setminus \{ {X}^0{Z}^0 \}$ of cardinality $p^2 - 1$ can be partitioned into $p+1$ subsets containing each $p-1$ commuting matrices.  

As an example, let us consider the case $d=5$. For this case, we are left with the six following sets of four commuting matrices
           \begin{eqnarray*}  	   
&{\cal V}_0      =  \{  01 ,  02 ,  03 ,  04  \}, \quad	   	   
 {\cal V}_1      =  \{  10 ,  20 ,  30 ,  40  \}, \quad 
 {\cal V}_2      =  \{  11 ,  22 ,  33 ,  44  \}& \\
&{\cal V}_3      =  \{  12 ,  24 ,  31 ,  43  \}, \quad
 {\cal V}_4      =  \{  13 ,  21 ,  34 ,  42  \}, \quad
 {\cal V}_5      =  \{  14 ,  23 ,  32 ,  41  \}& 
          \end{eqnarray*} 
where $ab$ is used as an abbreviation of ${X}^a {Z}^b$. 

More generally, for $d=p$ with $p$ prime, the $p+1$ sets of $p-1$ commuting matrices 
are easily seen to be                 
           \begin{eqnarray*}  
{\cal V}_0       &=      &  \{ {X}^0 {Z}^a        \ : \  a = 1, 2, \cdots, p-1 \} 
        \\                   
{\cal V}_1       &=      &  \{ {X}^a {Z}^0        \ : \  a = 1, 2, \cdots, p-1 \}   
        \\
{\cal V}_2       &=      &  \{ {X}^a {Z}^a        \ : \  a = 1, 2, \cdots, p-1 \} 
        \\
{\cal V}_3       &=      &  \{ {X}^a {Z}^{2a}     \ : \  a = 1, 2, \cdots, p-1 \} 
              \\
                 &\vdots  & 
        \\
 {\cal V}_{p-1}  &=      &  \{ {X}^a {Z}^{(p-2)a} \ : \  a = 1, 2, \cdots, p-1 \} 
        \\  
 {\cal V}_{p}    &=      &  \{ {X}^a {Z}^{(p-1)a} \ : \  a = 1, 2, \cdots, p-1 \}       
           \end{eqnarray*} 
Each of the $p+1$ sets ${\cal V}_0, {\cal V}_1, \cdots, {\cal V}_{p}$ can be put in a one-to-one correspondence 
with one basis of the complete set of $p+1$ MUBs. In fact, ${\cal V}_0$ is associated with the computational basis 
while ${\cal V}_1, {\cal V}_2, \cdots, {\cal V}_{p}$ are associated with the $p$ remaining MUBs in view of 
           \begin{eqnarray*} 
{V_{a}} \in {\cal V}_{a + 1} = \{ {X}^b {Z}^{ab} \ : \ b = 1, 2, \cdots, p-1 \}, \quad a = 0, 1, \cdots, p-1
           \end{eqnarray*} 
More precisely, we have 
$$
Z \in {\cal V}_{0}, \quad X \in {\cal V}_{1}, \quad X Z \in {\cal V}_{2}, \quad \cdots, \quad X {Z}^{p-1} \in {\cal V}_{p}  
$$
The eigenvectors of the $p+1$ unitary operators
$$
Z, \quad X, \quad X Z, \quad \cdots, \quad X {Z}^{p-1} 
$$
generate $p+1$ MUBs (one basis is associated with each of the $p+1$ operators). 

\subsection{Galois field approach to MUBs} \label{Galois field approach}

The existence of a complete set of $p^m + 1$ MUBS in $\mathbb{C}^{p^m}$ ($p$ prime and $m$ positive integer) is an indication of a possible utility of Galois fields and Galois rings for the construction of MUBs in $\mathbb{C}^{p^m}$ ($p$ prime, $m \geq 2$). Indeed, the passage from the case $d = p$ to the case $d = p^m$ ($p$ prime, $m \geq 2$) can be achieved by considering the Galois field $\mathbb{GF}(p^m)$ for $p$ odd prime and the Galois ring $\mathbb{GR}(2^2 , m)$ for $p=2$ \cite{19Klappenecker,02Kibler}. In this section, we shall deal with the construction of a complete set of $p^m + 1$ MUBs in $\mathbb{C}^{p^m}$, corresponding to the case of $m$ qupits, via the use of the Galois field $\mathbb{GF}(p^m)$ for $p$ odd prime and $m$ greater than 1. 

\subsubsection{The computational basis}

We first have to define the computational basis $B_{p^m}$ in the framework of $\mathbb{GF}(p^m)$, $p$ odd prime and $m \geq 2$. The vectors of the basis $B_{p^m}$ of the Hilbert space $\mathbb{C}^{p^m}$ can be labeled by the elements $x$ of the Galois field $\mathbb{GF}(p^m)$. This can be done in two ways according to as the elements $x$ are taken in the monomial form ($x = 0, \> \alpha^{\ell}$ with $\ell = 1, 2, \cdots, p^m - 1$) or in the polynomial form 
($x = [x_0 x_1 \cdots x_{m-1}]$ with $x_0, x_1, \cdots, x_{m-1} \in \mathbb{F}_p$). In both cases, we have 
$$
B_{p^m} = \{ | 0       \rangle \ {\rm or} \ \phi_0, \quad 
             | 1       \rangle \ {\rm or} \ \phi_1, \quad \cdots, \quad 
						 | p^m - 1 \rangle \ {\rm or} \ \phi_{p^m - 1} \}
$$
in terms of vectors or column vectors. More precisely, this can be achieved as follows. 
\begin{itemize}
	\item 
In the monomial form, we define the vectors of $B_{p^m}$ via the correspondences 
$$
x = 0 \mapsto | 0 \rangle \ {\rm or} \ \phi_0, \quad x = \alpha^{\ell} \mapsto | \ell \rangle \ {\rm or} \ \phi_{\ell} \ {\rm with} \ \ell = 1, 2, \cdots, p^m - 1
$$
where $\alpha$ is a primitive element of $\mathbb{GF}(p^m)$. 
	\item
In the polynomial form, we can range the vectors of $B_{p^m}$ in the order $0, 1, \cdots, p^m - 1$ by adopting the lexicographical order for the elements 
$[x_0 x_1 \cdots x_{m-1}]$. 
\end{itemize}
These notations are reminiscent of those employed for the computational basis 
$$
B_p = \{ | 0   \rangle \ {\rm or} \ \phi_0, \quad  
         | 1   \rangle \ {\rm or} \ \phi_1, \quad  
				 \cdots, \quad 
				 | p-1 \rangle \ {\rm or} \ \phi_{p-1} \}
$$ 
corresponding to the limit case $m = 1$.

\subsubsection{Shift and phase operators for $\mathbb{GF}(p^m)$} \label{Weyl pair for Galois field}

The notion of Weyl pair can be extended to any Galois field $\mathbb{GF}(p^m)$ with $p$ (even or odd) prime and $m \geq 2$. Let $x$ and $y$ be two elements of $\mathbb{GF}(p^m)$ and $\phi_y$ be the basis column vector of $B_{p^m}$ associated with $y$. For fixed $x$, we define the matrices 
$\hat{X}_x$ (shift operators) and $\hat{Z}_x$ (phase operators) via the actions 
$$
\hat{X}_x \phi_y = \phi_{y - x}, \quad 
\hat{Z}_x \phi_y = \chi(xy) \phi_{y} = {\rm e}^{{\rm i} \frac{2 \pi}{p} {\rm Tr}(xy)} \phi_{y}
$$
where $y$ is arbitrary. One easily verifies the properties
$$
\hat{X}_{x + y} = \hat{X}_x \hat{X}_y = \hat{X}_y \hat{X}_x, \quad 
\hat{Z}_{x + y} = \hat{Z}_x \hat{Z}_y = \hat{Z}_y \hat{Z}_x
$$
and 
$$
\hat{X}_x \hat{Z}_y - \chi(xy) \hat{Z}_y \hat{X}_x = O_{p^m}, \quad \chi(xy) = {\rm e}^{{\rm i} \frac{2 \pi}{p} {\rm Tr}(xy)}
$$
In the limit case $m = 1$ (i.e.~for the base field $\mathbb{F}_p$) the matrices 
$$
X = \hat{X}_1, \quad Z = \hat{Z}_1
$$
corresponding to $x = y = 1$ satisfy 
$$
X Z - {\rm e}^{{\rm i} \frac{2 \pi}{p}} Z X = O_p
$$
to be compared with the relations satisfied by the Weyl pair $(X , Z)$ defined in \ref{Weylpair}.  

\subsubsection{Bases in the frame of $\mathbb{GF}(p^m)$}

We might use the Weyl pair $(X_x,Z_y)$ defined in the framework of $\mathbb{GF}(p^m)$, see Section \ref{Weyl pair for Galois field}, for determining a complete set of $p^{m} + 1$ MUBs in $\mathbb{C}^{p^m}$ in a way similar to that used for $m = 1$ with the help of the matrix $V_a$ for $a$ in 
$\mathbb{F}_p$. However, it is quicker to start from the alternative formula (\ref{alternativef}) giving MUBs in $\mathbb{C}^{p}$ in order to generate a formula for $\mathbb{C}^{p^m}$ giving back Eq.~(\ref{alternativef}) in $\mathbb{C}^{p}$ in the limit case $m=1$. In this direction, a possible way to pass from the basis vector 
$$
 \frac{1}{\sqrt{p}} \sum_{x \in \mathbb{F}_p} {\rm e}^{{\rm i} \frac{2 \pi}{p}(a x + \alpha) x} | x \rangle
$$
of $\mathbb{C}^{p}$ to a basis vector of $\mathbb{C}^{p^m}$ is to replace
$$
{\rm e}^{{\rm i} \frac{2 \pi}{p}(a x + \alpha) x}, \quad a, \alpha, x \in \mathbb{F}_p
$$
by 
$$
\chi(a x^2 + \alpha x) = {\rm e}^{{\rm i} \frac{2 \pi}{p}{\rm Tr}(a x^2 + \alpha x)}, \quad a, \alpha, x \in \mathbb{GF}(p^m)
$$
where $\chi$ is the canonical additive character of $\mathbb{GF}(p^m)$. This yields the two following propositions. 

\medskip
\noindent $\blacktriangleright$~{\bf Proposition.} For $p$ odd prime and $m \geq 2$, the set 
$$
B_a = \{ | a \alpha \rangle \ : \ \alpha \in \mathbb{GF}(p^m) \}
$$
where 
$$
| a \alpha \rangle = \frac{1}{\sqrt{p^m}} \sum_{x \in \mathbb{GF}(p^m)} {\rm e}^{{\rm i} \frac{2 \pi}{p} {\rm Tr}(a x^2 + \alpha x) } | x \rangle, \quad 
a \in \mathbb{GF}(p^m)
$$
constitutes an orthonormal basis of $\mathbb{C}^{p^m}$.~$\blacktriangleleft$ 

\medskip 
{\em Proof}. See the proof of the next proposition.     \hfill $\square$ 

\medskip
Note that for $m=1$
$$
{\rm Tr}( a x^2 + \alpha x ) = a x^2 + \alpha x
$$
so that the vector $| a \alpha \rangle$ coincides with the vector $| a \alpha \rangle'$ derived in Section \ref{alternative formula}. This explains why we chose to extend the alternative formula (\ref{alternativef}) valid for $\mathbb{C}^p$ to the case $\mathbb{C}^{p^m}$. Indeed, the same kind of extension applied to the formula (\ref{masterformula}) is not possible since ${\rm Tr}[\frac{1}{2} n (p - n) a + n \alpha]$ does not make sense. 

\subsubsection{MUBs in the frame of $\mathbb{GF}(p^m)$}

$\blacktriangleright$~{\bf Proposition.} For $p$ odd prime and $m \geq 2$, the $p^m$ bases $B_a$, $a$ ranging in $\mathbb{GF}(p^m)$, constitute with the computational basis $B_{p^m}$ a complete set of $p^m + 1$ MUBs in $\mathbb{C}^{p^m}$.~$\blacktriangleleft$ 

\medskip 
{\em Proof}. Let $| a \alpha \rangle$ and $| b \beta \rangle$ two vectors belonging to the bases $B_a$ and $B_b$, respectively. We have 
$$
\langle a \alpha | b \beta \rangle  = \frac{1}{p^m}
\sum_{x \in \mathbb{GF}(p^m)} {\rm e}^{ {\rm i} \frac{2 \pi}{p} {\rm Tr}[ (b-a) x^2 + (\beta - \alpha) x ] }, \quad 
a, b, \alpha, \beta \in \mathbb{GF}(p^m)
$$ 
By using \cite{Weil,Berndt12,02Kibler}
$$
\left| \sum_{x \in \mathbb{GF}(p^m)} {\rm e}^{ {\rm i} \frac{2 \pi}{p} {\rm Tr}( u x^2 + v x ) } \right| 
= \sqrt{p^m}, \quad u \in {\mathbb{GF}(p^m)}^*, \quad v \in \mathbb{GF}(p^m)
$$
(valid for $p$ odd prime), we obtain 
$$
| \langle a \alpha | b \beta \rangle | = \cases{ \delta_{\alpha , \beta} \ {\rm if} \ b = a \cr \cr \frac{1}{\sqrt{p^m}} \ {\rm if} \ b \not= a
}
$$ 
or in compact form 
\begin{eqnarray*}
| \langle  a \alpha |  b \beta \rangle | = \delta_{a,b} \delta_{\alpha,\beta} + \frac{1}{\sqrt{p^m}} (1 - \delta_{a,b}) 
\end{eqnarray*}
which shows that $B_a$ is an orthonormal basis and that the couple ($B_{a}, B_{b}$) with $b \not= a$ is a couple of unbiased bases. Of course, each basis $B_a$ is unbiased to the computational basis $B_{p^m}$. We thus end up with a total of $p^m + 1$ MUBs as desired.              \hfill $\square$ 

\medskip
The previous result applies in the limit case $m=1$ for which we recover the $p+1$ MUBs in $\mathbb{C}^p$. 

\subsection{Galois ring approach to MUBs} \label{Galois ring approach}

In dimension $d = 2^m$, $m \geq 2$, the use of the Galois field $\mathbb{GF}(2^m)$ for constructing a complete set of $2^m + 1$ MUBs in $\mathbb{C}^{2^m}$ according to the method employed in Section \ref{Galois field approach} for $d = p^m$, $p$ odd prime, would lead to a no-win situation because ${\rm gcd}(2 , 2^m) \not= 1$ 
(while ${\rm gcd}(2 , p^m) = 1$ for $p$ odd prime). For $d = 2^m$, which corresponds to the case of $m$ qubits, we can use the Galois ring $\mathbb{GR}(2^2 , m)$, denoted $R_{4^m}$ too, for constructing MUBs in $\mathbb{C}^{2^m}$. 

\subsubsection{Bases in the frame of $\mathbb{GR}(2^2 , m)$} \label{MUBs en anneau}

We start with the residue class ring 
$$
\mathbb{GR}(2^2, m) = \mathbb{Z}_{2^2}[\xi] / \langle P_m(\xi) \rangle
$$ 
where $P_m(x)$ is a monic basic irreducible polynomial of degree $m$ (i.e.~its restriction $\overline{P_m(x)} = P_m(x)$ modulo 2 is irreducible over 
$\mathbb{Z}_{2}$). The $2^m$ vectors of the computational basis $B_{2^m}$ are labeled by the $2^m$ elements of the Teichm\"uller set $T_m$ associated with the ring $\mathbb{Z}_{2^2}[\xi] / \langle P_m(\xi) \rangle$. Thus 
$$
B_{2^m} = \{ | x \rangle \ : \ x \in T_m\}
$$
(the set $T_m$ and the ring $\mathbb{GR}(2^2, m)$ contain $2^m$ and $4^m$ elements, respectively). 

\medskip
\noindent $\blacktriangleright$~{\bf Proposition.} For $a$ and $\alpha$ in $T_m$, let  
\begin{eqnarray*}
| a \alpha \rangle 
= \frac{1}{\sqrt{2^m}} \sum_{x \in T_m} \chi[(a + 2 \alpha) x] | x \rangle
= \frac{1}{\sqrt{2^m}} \sum_{x \in T_m} {\rm e}^{{\rm i} \frac{2 \pi}{4} {\rm Tr}(a x + 2 \alpha x)} | x \rangle 
= \frac{1}{\sqrt{2^m}} \sum_{x \in T_m} {\rm i}^{{\rm Tr}(a x + 2 \alpha x)} | x \rangle 
\end{eqnarray*}
where $\chi$ is an additive character vector of $\mathbb{GR}(2^2, m)$ and the trace takes its values in 
$\mathbb{Z}_4$. For fixed $a$ in $T_m$, the set 
$$
B_a = \{ | a \alpha \rangle \ : \ \alpha \in T_m \}
$$
constitutes an orthonormal basis of $\mathbb{C}^{2^m}$.~$\blacktriangleleft$ 

\medskip 
{\em Proof}. See the proof of the next proposition.     \hfill $\square$ 

\medskip
Note that for $m = 1$
$$
{\rm Tr}( a x + 2 \alpha x ) = a x + 2 \alpha x
$$
so that  
$$
| a \alpha \rangle = \frac{1}{\sqrt{2}} \sum_{x \in \mathbb{F}_2} {\rm i}^{a x + 2 \alpha x} | x \rangle  
\eqno (18)
$$
to be compared with the vector 
$$
|a \alpha \rangle =  
\frac{1}{\sqrt{2}} \sum_{x \in \mathbb{F}_2}
{\rm e}^{{\rm i} \frac{2 \pi}{2} [\frac{1}{2} a x (2 - x) + \alpha x]} \vert 1 - x \rangle = 
\frac{1}{\sqrt{2}} \sum_{x \in \mathbb{F}_2}
{\rm i}^{a x (2 - x) + 2 \alpha x} \vert 1 - x \rangle 
\eqno (19)
$$
given by the formula (\ref{masterformula}). In view of the fact that
$$
{\rm i}^{a x + 2 \alpha x} = 
{\rm i}^{a x (2 - x) + 2 \alpha x}
$$
for $x=0$ and $x=1$, the two vectors $| a \alpha \rangle$ in Eqs.~(18) and (19) are the same up to an interchange of the vectors 
$| 0 \rangle$ and $| 1 \rangle$. 

\subsubsection{MUBs in the frame of $\mathbb{GR}(2^2 , m)$}

$\blacktriangleright$~{\bf Proposition.} The $2^m$ bases $B_a$, with $m \geq 2$ and $a$ ranging in the Teich\"uller set $T_m$ associated with the Galois ring 
$\mathbb{GR}(2^2 , m)$, constitute with the computational basis $B_{2^m}$ a complete set of $2^m + 1$ MUBs in $\mathbb{C}^{2^m}$.~$\blacktriangleleft$ 

\medskip 
{\em Proof}. Let $| a \alpha \rangle$ and $| b \beta \rangle$ two vectors belonging to the bases $B_a$ and $B_b$, respectively. We have 
\begin{eqnarray*}
\langle a \alpha | b \beta \rangle 
= \frac{1}{2^m} \sum_{x \in T_m} {\rm e}^{{\rm i} \frac{\pi}{2}{\rm Tr}[(b - a + 2 \beta - 2 \alpha) x]} 
\end{eqnarray*}
By using \cite{Weil,Berndt12,02Kibler}
$$
 \left| \sum_{x \in T_m} {\rm e}^{{\rm i} \frac{\pi}{2}{\rm Tr}(u x)} \right|  
= \cases{0   \ {\rm if} \ u \in 2 T_m, \ u \not= 0 \cr \cr 
         2^m \ {\rm if} \ u = 0 \cr \cr
				 \sqrt{2^m} \ {\rm otherwise} 
}
$$
we obtain 
$$
| \langle a \alpha | b \beta \rangle | = \cases{ \delta_{\alpha , \beta} \ {\rm if} \ b = a \cr \cr \frac{1}{\sqrt{2^m}} \ {\rm if} \ b \not= a
}
$$ 
or in compact form 
\begin{eqnarray*}
| \langle  a \alpha |  b \beta \rangle | = \delta_{a,b} \delta_{\alpha,\beta} + \frac{1}{\sqrt{2^m}} (1 - \delta_{a,b}) 
\end{eqnarray*}
which shows that $B_a$ is an orthonormal basis and that the couple ($B_{a}, B_{b}$) with $b \not= a$ is a couple of unbiased bases. Of course, each basis $B_a$ is unbiased to the computational basis $B_{2^m}$. We thus end up with a total of $2^m + 1$ MUBs and we are done.             \hfill $\square$ 

\medskip
The previous result applies in the limit case $m=1$ for which we can recover the $2+1$ MUBs in $\mathbb{C}^2$. 

\subsubsection{One-qubit system}

For $m = 1$, the $2^m = 2$ vectors of the computational basis $B_{2}$ are labeled with the help of the two elements of the Teichm\"uller set $T_1 = \mathbb{Z}_2$ of the Galois ring $\mathbb{GR}(2^2 , 1) = \mathbb{Z}_{2^2}$. Thus, the basis $B_{2}$ is 
     \begin{eqnarray*}
B_{2} &:& | 0 \rangle = \pmatrix{ 
1 \cr
0 \cr
}, \quad | 1 \rangle = \pmatrix{ 
0 \cr
1 \cr
}
		 \end{eqnarray*}
The vectors $| a \alpha \rangle$ of the basis $B_a$ ($a \in T_1$) are given by (see \ref{MUBs en anneau})
$$
| a \alpha \rangle = \frac{1}{\sqrt{2}} \sum_{x = 0}^{1} {\rm i}^{(a + 2 \alpha) x} | x \rangle, \quad 
\alpha \in T_1 = \{ 0,1 \}
$$
This yields the two unbiased bases
\begin{eqnarray*}
 B_0 &:&  | 0 0 \rangle = \frac{| 0 \rangle +         | 1 \rangle}{\sqrt{2}}, \quad 
          | 0 1 \rangle = \frac{| 0 \rangle -         | 1 \rangle}{\sqrt{2}}  \\
 B_1 &:&  | 1 0 \rangle = \frac{| 0 \rangle + {\rm i} | 1 \rangle}{\sqrt{2}}, \quad 
          | 1 1 \rangle = \frac{| 0 \rangle - {\rm i} | 1 \rangle}{\sqrt{2}}    				
\end{eqnarray*} 
which, together with the computational basis $B_2$, form a complete set of $2+1 = 3$ MUBs in $\mathbb{C}^2$. Note that the bases $B_0$ and $B_1$ are in agreement (up to phase factors and a rearrangement of the vectors inside $B_1$) with the bases $B_0$ and $B_1$ derived in Section 
\ref{complete-set-of-MUBs}.

\subsubsection{Two-qubit system}

For $m = 2$, the $2^m = 4$ vectors of the computational basis $B_{4}$ are labeled with the help of the four elements of the Teichm\"uller set 
$T_2 = \{ 0, \ \beta^1, \ \beta^2 = 3 + 3 \beta, \ \beta^3 = 1 \}$ of the Galois ring $\mathbb{GR}(2^2 , 2)$ (here, we use $\beta$ instead of $\alpha$ in order to avoid confusion with the index $\alpha$ in $| a \alpha \rangle$). Thus, the basis $B_{4}$ is 
     \begin{eqnarray*}
B_{4} &:& | 0 \rangle = \pmatrix{ 
1 \cr
0 \cr
0 \cr
0 \cr
}, \ | \beta^1 \, {\rm or} \, 1 \rangle = \pmatrix{ 
0 \cr
1 \cr
0 \cr
0 \cr
}, \ | \beta^2 \, {\rm or} \, 2 \rangle = \pmatrix{ 
0 \cr
0 \cr
1 \cr
0 \cr
}, \ | \beta^3 \, {\rm or} \, 3 \rangle = \pmatrix{ 
0 \cr
0 \cr
0 \cr
1 \cr
}
		 \end{eqnarray*}
The vectors $| a \alpha \rangle$ of the basis $B_a$ 
($a = 0, \ \beta^1 \ {\rm or} \ 1,  
         \ \beta^2 \ {\rm or} \ 2,
         \ \beta^3 \ {\rm or} \ 3$) 
are given by (see \ref{MUBs en anneau})
$$
| a \alpha \rangle = \frac{1}{2} \sum_{x \in T_2} {\rm i}^{{\rm Tr}(ax + 2 \alpha x)} | x \rangle, \quad 
\alpha \in T_2 = \{ 0, \ \beta^1, \ \beta^2 = 3 + 3 \beta, \ \beta^3 = 1 \}
$$
with 
$$
{\rm Tr}(ax + 2 \alpha x) = ax + 2 \alpha x + \phi(ax + 2 \alpha x)
$$
where $\phi$ is the generalized Frobenius map $\mathbb{GR}(2^2 , 2) \rightarrow \mathbb{GR}(2^2 , 2)$. The correspondence between the indexes $a, \alpha$ in $| a \alpha \rangle$ and the elements 
$0, \beta^1,  \beta^2, \beta^3$ of $T_2$ is as follows 
\begin{eqnarray*}
0         \leftrightarrow a \ {\rm or} \ \alpha = 0, \ 
\beta^1   \leftrightarrow a \ {\rm or} \ \alpha = 1, \ 
\beta^2   \leftrightarrow a \ {\rm or} \ \alpha = 2, \ 
\beta^3   \leftrightarrow a \ {\rm or} \ \alpha = 3
\end{eqnarray*}
This yields the four unbiased bases
\begin{eqnarray*}
B_0 &:&  
| 00 \rangle = \frac{1}{2} \pmatrix{
1 \cr
1 \cr
1 \cr
1 \cr
}, \,
| 01 \rangle = \frac{1}{2} \pmatrix{
1 \cr
-1 \cr
1 \cr
-1 \cr
}, \,
| 02 \rangle = \frac{1}{2} \pmatrix{
1 \cr
1 \cr
-1 \cr
-1 \cr
}, \,
| 03 \rangle = \frac{1}{2} \pmatrix{
1 \cr
-1 \cr
-1 \cr
1 \cr
} 
\\
B_1 &:& 
| 12 \rangle = \frac{1}{2} \pmatrix{
1 \cr
-{\rm i} \cr
1 \cr
{\rm i} \cr
}, \,
| 11 \rangle = \frac{1}{2} \pmatrix{
1 \cr
{\rm i} \cr
-1 \cr
{\rm i} \cr
}, \,
| 13 \rangle = \frac{1}{2} \pmatrix{
1 \cr
{\rm i} \cr
1 \cr
-{\rm i} \cr
}, \,
| 10 \rangle = \frac{1}{2} \pmatrix{
1 \cr
-{\rm i} \cr
-1 \cr
-{\rm i} \cr
}
\\
B_2 &:& 
| 21 \rangle = \frac{1}{2} \pmatrix{
1 \cr
1 \cr
-{\rm i} \cr
{\rm i} \cr
}, \,
| 22 \rangle = \frac{1}{2} \pmatrix{
1 \cr
-1 \cr
{\rm i} \cr
{\rm i} \cr
}, \,
| 20 \rangle = \frac{1}{2} \pmatrix{
1 \cr
-1 \cr
-{\rm i} \cr
-{\rm i} \cr
}, \,
| 23 \rangle = \frac{1}{2} \pmatrix{
1 \cr
1 \cr
{\rm i} \cr
-{\rm i} \cr
}
\\
B_3 &:&  
| 33 \rangle = \frac{1}{2} \pmatrix{
1 \cr
{\rm i} \cr
{\rm i} \cr
-1 \cr
}, \,
| 32 \rangle = \frac{1}{2} \pmatrix{
1 \cr
-{\rm i} \cr
{\rm i} \cr
1 \cr
}, \,
| 31 \rangle = \frac{1}{2} \pmatrix{
1 \cr
{\rm i} \cr
-{\rm i} \cr
1 \cr
}, \,
| 30 \rangle = \frac{1}{2} \pmatrix{
1 \cr
-{\rm i} \cr
-{\rm i} \cr
-1 \cr
} 
\end{eqnarray*}  
We thus end up with $4+1 = 5$ bases ($B_0$ to $B_4$) which form a complete set of MUBs in $\mathbb{C}^4$. Note that the bases $B_0$, $B_1$, $B_2$ and 
$B_3$ coincide with the bases $W_{00}$, $W_{10}$, $W_{01}$ and $W_{11}$ derived from tensor products, respectively; for the purpose of comparison, the vectors $| a \alpha \rangle$ are listed in the same order for each of the couples $(B_0, W_{00})$, $(B_1, W_{10})$, $(B_2, W_{01})$ and 
$(B_3, W_{11})$, see \ref{MUBs by tensor product}.

\section{Closing remarks}

During the last two decades, quantum information and quantum computing have been the object of considerable progresses both in theoretical and experimental physical sciences, scientific engineering, discrete mathematics and quantum informatics. In the present days, there exit several quantum computer languages and, although quantum devices are mainly developed in academic and private laboratories, the scientific community has access to some quantum computers (e.g., access to the 5-qubit quantum computer of the IBM Quantum Experience \cite{IBM56}) and to quantum simulators (e.g., access to the 41-qubit ATOS Quantum Learning Machine \cite{ATOS57}). In the medium term, the accent shall be put, among others, on 
(i) the development of new quantum algorithms that outperform classical ones, 
(ii) the production of qubits robust to decoherence, 
(iii) the increase of the lifetime of quantum memories, 
(iv) the development of quantum networks working over a few thousands of kilometres 
(v) the realization of 50-100 qubit computers, and 
(vi) the test of quantum supremacy. Long way before the realization of a universal quantum computer!

From the side of the mathematical aspects of MUBs, some further developments and a few open problems should be mentioned. It would be interesting to see if Cayley-Dickson algebras of dimension $d=2^N$ could be used for providing a geometrical approach to entanglement of $N$ qubits with $N > 3$. Furthermore, the problem of the determination of the maximum number $N(d)$ of MUBs in composite dimension $d$ is still an unsolved problem (except in the case where $d$ is a power of a prime number). The two conjectures listed in Section \ref{conjectures} do not very much help, probably because they lead to two equivalent problems for which the solutions are as difficult to find as those of the initial problem. As far as the second conjecture is concerned, the recent work \cite{Sriwongsa} on orthogonal decompositions of sl$(n,R)$ over a finite commutative ring with identity $R$ is very appealing. Finally, even in the simplest case where $d=6$, the maximum number $N(6)$ of MUBs is not known (to the best of the author knowledge). However, for $d = 6$ there are numerous numerical evidences that $N(6) = 3$ \cite{13Zauner,25Grassl,36Bengtsson,38Butterley,Brierley,45Brierley,McNulty,50McNulty}. The number $N(6) = 3$ is equal indeed to the number of {\em weak} mutually unbiased bases associated with the smallest prime divisor of 6 (the recently introduced notion of weak MUBs in dimension $d$ corresponds to the definition (\ref{definition des mubs}) where $\sqrt{d}$ is replaced by $\sqrt{f}$ where $f$ is a prime divisor of $d$ \cite{49Shalaby,54Olupitan}).

\section*{Acknowledgements}

This paper was presented at the 20th International Workshop on Computer Algebra in Scientific Computing (CASC 2018). The author wishes to thank  
Vladimir P. Gerdt (Dubna) for his kind invitation to give an invited talk at CASC 2018 and Andreas Weber (Bonn) for his encouragement to put the text of the talk in a form convenient for a community of computer engineers and mathematicians. He is also indebted to Wolfram Koepf (Kassel) and Fran\c cois Boulier (Lille) for their logistic help during the preparation of this paper.


\newpage


\begin{thebibliography}{99}

\small{

			\bibitem{01Nielsen} M.A. Nielsen and I.L. Chuang, 
Quantum Computation and Quantum Information, Cambridge University Press, Cambridge (2003). 


			\bibitem{Weyl} H. Weyl, 
The Theory of Groups and Quantum Mechanics (Dover Publications, New York, 1931). 


			\bibitem{Weil} A. Weil, 
On some exponential sums, {\em Proc. Nat. Acad. Sci. USA} 34 (1948) 204-207. 


			\bibitem{03Schwinger} J. Schwinger, 
Unitary operator bases, {\em Proc. Nat. Acad. Sci. USA} 46 (1960) 570-579. 


			\bibitem{04Ivanovic} I.D. Ivanovi\'c, 
Geometrical description of quantal state determination, {\em J. Phys. A: Math. Gen.} 14 (1981) 3241-3245. 


			\bibitem{Kostrikin} A.I. Kostrikin, I.A. Kostrikin and V.A. Ufnarovski\u{\i}, 
Orthogonal decompositions of simple Lie algebras (type $A_n$), {\em Trudy Mat. Inst. Steklov} 158 (1981) 105-120. (in Russian)


			\bibitem{05Wootters} W.K. Wootters and W.H. Zurek, 
A single quantum cannot be cloned, {\em Nature} 299 (1982) 802. 


			\bibitem{06Bennett} C.H. Bennett and G. Brassard, 
Quantum cryptography: Public key distribution and coin tossing, {\em Proc. IEEE Int. Conf. Computers, Systems, and Signal Processing, Bangalore, India} (1984) 175-179. 


			\bibitem{07Wootters} W.K. Wootters, 
Quantum mechanics without probability amplitudes, {\em Found. Phys.} 16 (1986) 391-405. 


			\bibitem{Wootters1987} W.K. Wootters, 
A Wigner function formulation of finite-state quantum mechanics, {\em Ann. Phys. (N.Y.)} 176 (1987) 1-21. 


			\bibitem{08Patera} J. Patera and H. Zassenhaus, 
The Pauli matrices in $n$ dimensions and finest gradings of simple Lie algebras of type $A_{n-1}$, {\em J. Math. Phys.} 29 (1988) 665-673. 


			\bibitem{Lambert} D. Lambert and M. Kibler, 
An algebraic and geometric approach to non-bijective quadratic transformations, {\em J. Phys. A: Math. Gen.} 21 (1988) 307-343. 


			\bibitem{09Wootters} W.K. Wootters and B.D. Fields, 
Optimal state-determination by mutually unbiased measurements, {\em Ann. Phys. (N.Y.)} 191 (1989) 363-381. 


			\bibitem{10Bennett} C.H. Bennett, G. Brassard, C. Cr\'epeau, R. Jozsa, A. Peres, and W.K. Wootters, 
Teleporting an unknown quantum state via dual classical and Einstein-Podolsky-Rosen channels, {\em Phys. Rev. Lett.} 70 (1993) 1895. 


			\bibitem{Kostrikinlivre} A.I. Kostrikin and P.H. Tiep, 
Orthogonal decompositions and integral lattices (Walter de Gruyter, Berlin, 1994). 


			\bibitem{11Calderbank} A.R. Calderbank, P.J. Cameron, W.M. Kantor and J.J. Seidel, 
$\mathbb{Z}_4$--Kerdock codes, orthogonal spreads, and extremal Euclidean line-sets, {\em Proc. London Math. Soc.} 75 (1997) 436-480. 


			\bibitem{Berndt12} B.C. Berndt, R.J. Evans and K.S. Williams, Gauss and Jacobi sums (Wiley, New York, 1998).


			\bibitem{13Zauner} G. Zauner, 
Quantendesigns: Grundz\"uge einer nichtcommutativen Designtheorie, {\em Diploma Thesis}, University of Wien (1999). 


			\bibitem{14Englert} B.-G. Englert and Y. Aharonov, 
The mean king's problem: prime degrees of freedom, {\em Phys. Lett. A} 284 (2001) 1-5. 


			\bibitem{Mosseri} R. Mosseri and R. Dandoloff, 
Geometry of entangled states, Bloch spheres and Hopf fibrations, {\em J. Phys. A: Math. Gen.} 34 (2001) 10243-10252. 


			\bibitem{15Bandyopadhyay} S. Bandyopadhyay, P.O. Boykin, V. Roychowdhury and F. Vatan, 
A new proof for the existence of mutually unbiased bases, {\em Algorithmica} 34 (2002) 512-528. 


			\bibitem{16Lawrence} J. Lawrence, \v{C}. Brukner, and A. Zeilinger, 
Mutually unbiased binary observable sets on N qubits, {\em Phys. Rev. A} 65 (2002) 032320 (5pp). 


			\bibitem{17Chaturvedi} S. Chaturvedi, 
Aspects of mutually unbiased bases in odd prime power dimensions, {\em Phys. Rev. A} 65 (2002) 044301 (3pp). 


			\bibitem{18Cerf} N.J. Cerf, M. Bourennane, A. Karlsson and N. Gisin, 
Security of quantum key distribution using $d$-level systems, {\em Phys. Rev. Lett.} 88 (2002) 127902 (4pp). 


			\bibitem{Aravind} P.K. Aravind, 
Solution to the King's problem in prime power dimensions, {\em Z. Naturforsch.} 58a (2003) 85-92. 


			\bibitem{Lawrence} J. Lawrence, 
Mutually unbiased bases and trinary operator sets for N qutrits, {\em Phys. Rev. A} 70 (2004) 012302 (10pp). 


			\bibitem{19Klappenecker} A. Klappenecker and M. R\"otteler, 
Constructions of mutually unbiased bases, {\em Lect. Notes Comput. Sci.} 2948 (2004) 137-144. 


			\bibitem{20Gibbons} K.S. Gibbons, M.J. Hoffman and W.K. Wootters, 
Discrete phase space based on finite fields, {\em Phys. Rev. A} 70 (2004) 062101 (23pp). 


			\bibitem{21Pittenger} A.O. Pittenger and M.H. Rubin, 
Mutually unbiased bases, generalized spin matrices and separability, {\em Linear Algebr. Appl.} 390 (2004) 255-278. 


			\bibitem{Vourdas} A. Vourdas, 
Quantum systems with finite Hilbert space, {\em Rep. Prog. Phys.} 67 (2004) 267-320. 


			\bibitem{22Saniga} M. Saniga, M. Planat and H. Rosu, 
Mutually unbiased bases and finite projective planes, {\em J. Opt. B: Quantum Semiclassical Opt.} 6 (2004) L19-L20. 


			\bibitem{Hayashi} A. Hayashi, M. Horibe and T. Hashimoto, 
Mean king's problem with mutually unbiased bases and orthogonal Latin squares, {\em Phys. Rev. A} 71 (2005) 052331 (4pp).              


			\bibitem{Paz} J.P. Paz, A.J. Roncaglia and M. Saraceno, 
Qubits in phase space: Wigner-function approach to quantum-error correction and the mean-king problem, {\em Phys. Rev. A} 72 (2005) 012309 (19pp). 

			\bibitem{23Wocjan} P. Wocjan and T. Beth, 
New construction of mutually unbiased bases in square dimensions, {\em Quantum Inf. Comput.} 5 (2005) 93-101. 


			\bibitem{24Archer} C. Archer, 
There is no generalization of known formulas for mutually unbiased bases, {\em J. Math. Phys.} 46 (2005) 022106 (11pp). 


			\bibitem{25Grassl} M. Grassl, 
On SIC-POVMs and MUBs in dimension 6, in {\em Proceedings ERATO Conference on Quantum Information Science (EQIS'04), 
J. Gruska editor, Tokyo} (2005) 60-61 and arXiv:0406175v2 [quant-ph] (2009). 	


			\bibitem{Grassl} M. Grassl, 
Tomography of quantum states in small dimensions, {\em Elec. Notes Discrete Math.} 20 (2005) 151-164. 


			\bibitem{26Klappenecker} A. Klappenecker and M. R\"otteler, 
Mutually unbiased bases are complex projective 2-designs, in {\em Proceedings of the 2005 IEEE International Symposium on Information Theory (Adelaide, Australia)} (2005) 1740-1744. 


			\bibitem{27Bengtsson} I. Bengtsson and \AA. Ericsson, 
Mutually unbiased bases and the complementary polytope, {\em Open Syst. Inf. Dyn.} 12 (2005) 107-120. 


			\bibitem{28Durt} T. Durt, 
About mutually unbiased bases in even and odd prime power dimensions, {\em J. Phys. A: Math. Gen.} 38 (2005) 5267-5283. 


			\bibitem{Pittenger} A.O. Pittenger and M.H. Rubin, 
Wigner functions and separability for finite systems, {\em J. Phys. A: Math. Gen.} 38 (2005) 6005-6036. 


			\bibitem{Durt} T. Durt, 
About the Mean King's problem and discrete Wigner distributions, {\em Int. J. Mod. Phys. B} 20 (2006) 1742-1760. 


			\bibitem{KiblerQTAM} M.R. Kibler, 
Angular momentum and mutually unbiased bases, {\em Int. J. Mod. Phys. B} 20 (2006) 1792-1801. 


			\bibitem{29Kibler} M.R. Kibler and M. Planat, 
A SU(2) recipe for mutually unbiased bases, {\em Int. J. Mod. Phys. B} 20 (2006) 1802-1807. 


			\bibitem{31Vourdas} A. Vourdas, 
Galois quantum systems, irreducible polynomials and Riemann surfaces, {\em J. Math. Phys.} 47 (2006) 092104 (15pp). 


			\bibitem{Heath} R.W. Heath, T. Strohmer and A.J. Paulraj, On quasi-orthogonal signatures for CDMA systems, {\em IEEE Trans. Inf. Theory} 52 (2006) 1217-1226. 


			\bibitem{32Vourdas} A. Vourdas, 
Quantum systems in finite Hilbert space: Galois fields in quantum mechanics, {\em J. Phys. A: Math. Theor.} 40 (2007) R285-R331. 


			\bibitem{33Klimov} A.B. Klimov, J.L. Romero, G. Bj\"ork and L.L. S\'anchez-Soto, 
Geometrical approach to mutually unbiased bases, {\em J. Phys. A: Math. Theor.} 40 (2007) 3987-3998 and 40 (2007) 9177. 


			\bibitem{34Sulc} P. \v{S}ulc and J. Tolar, 
Group theoretical construction of mutually unbiased bases in Hilbert spaces of prime dimensions, {\em J. Phys. A: Math. Theor.} 40 (2007) 15099 (13pp). 


			\bibitem{35Aschbacher} M. Aschbacher, A.M. Childs and P. Wocjan, 
The limitations of nice mutually unbiased bases, {\em J. Algebr. Comb.} 25 (2007) 111-123. 


			\bibitem{36Bengtsson} I. Bengtsson, W. Bruzda, \AA. Ericsson, J.-\AA. Larsson, W. Tadej and K. \.{Z}yczkowski, 
Mutually unbiased bases and Hadamard matrices of order six, {\em J. Math. Phys.} 48 (2007) 052106 (21pp). 


			\bibitem{37Boykin} P.O. Boykin, M. Sitharam, P.H. Tiep and P. Wocjan, 
Mutually unbiased bases and orthogonal decompositions of Lie algebras, {\em Quantum Inf. Comput.} 7 (2007) 371-382. 


			\bibitem{38Butterley} P. Butterley and W. Hall, 
Numerical evidence for the maximum number of mutually unbiased bases in dimension six, {\em Phys. Lett. A} 369 (2007) 5-8. 


			\bibitem{39Bj} G. Bj\"ork, J.L. Romero, A.B. Klimov and L.L. S\'anchez-Soto, 
Mutually unbiased bases and discrete Wigner functions, {\em J. Opt. Soc. Am. B} 24 (2007) 371-378. 


			\bibitem{40Klimov} A.B. Klimov, C. Mu\~noz, A. Fern\'andez and C. Saavedra, 
Optimal quantum-state reconstruction for cold trapped ions, {\em Phys. Rev. A} 77 (2008) 060303(R) (4pp). 


			\bibitem{Svetlichny} G. Svetlichny, 
Feynman's integral is about mutually unbiased bases, arXiv:0708.3079v3 [quant-ph] (2008).  


			\bibitem{Kibler} M.R. Kibler, 
Variations on a theme of Heisenberg, Pauli and Weyl, {\em J. Phys. A: Math. Theor.} 41 (2008) 375302 (19pp). 


			\bibitem{Brierley} S. Brierley and S. Weigert, 
Maximal sets of mutually unbiased quantum states in dimension six, {\em Phys. Rev. A} 78 (2008) 042312 (8pp). 


			\bibitem{45Brierley} S. Brierley and S. Weigert, 
Constructing mutually unbiased bases in dimension six, {\em Phys. Rev. A} 79 (2009) 052316 (13pp). 


			\bibitem{41Appleby} D.M. Appleby,  
SIC-POVMS and MUBS: geometrical relationships in prime dimension, in {\em AIP Conference Proceedings, 
Foundations of Probability and Physics-5 (V\"axj\"o, Sweden, 2008)} 1101 (2009) 223-232. 


			\bibitem{42Albouy} O. Albouy, 
The isotropic lines of $\mathbb{Z}_d^2$, {\em J. Phys. A: Math. Theor.} 42 (2009) 072001 (8pp). 


			\bibitem{43Tolar} J. Tolar and G. Chadzitaskos, 
Feynman's path integral and mutually unbiased bases, {\em J. Phys. A: Math. Theor.} 42 (2009) 245306 (11pp). 


			\bibitem{44Kibler} M.R. Kibler, 
An angular momentum approach to quadratic Fourier transform, Hadamard matrices, Gauss sums, mutually unbiased bases, the unitary group and the Pauli group, {\em J. Phys. A: Math. Theor.} 42 (2009) 353001 (28pp). 


			\bibitem{46Durt} T. Durt, B.-G. Englert, I. Bengtsson and K. \.{Z}yczkowski, 
On mutually unbiased bases, {\em Int. J. Quantum Inf.} 8 (2010) 535-640. 


			\bibitem{Dita} P. Di\c t\u a, 
Hadamard matrices from mutually unbiased bases, {\em J. Math. Phys.} 51 (2010) 072202 (20pp). 


			\bibitem{Zauner} G. Zauner, 
Quantum designs: Foundations of a noncommutative design theory, {\em Int. J. Quantum Inf.} 9 (2011) 445-507.


			\bibitem{48Daoud} M. Daoud and M.R. Kibler,  
Phase operators, phase states and vector phase states for $SU_3$ and $SU_{2,1}$, {\em J. Math. Phys.} 52 (2011)	082101 (21pp). 


			\bibitem{49Shalaby} M. Shalaby and A. Vourdas, 
Weak mutually unbiased bases, {\em J. Phys. A: Math. Theor.} 45 (2012) 052001 (15pp). 


			\bibitem{McNulty} D. McNulty and S. Weigert, 
The limited role of mutually unbiased product bases in dimension six, {\em J. Phys. A: Math. Theor.} 45 (2012) 102001 (5pp). 


			\bibitem{50McNulty} D. McNulty and S. Weigert,
All mutually unbiased product bases in dimension six, {\em J. Phys. A: Math. Theor.} 45 (2012) 135307 (22pp). 


			\bibitem{51Ghiu} I. Ghiu, 
Generation of all sets of mutually unbiased bases for three-qubit systems, {\em Phys. Scr.} T153 (2013) 014027 (5pp). 


			\bibitem{52Goyeneche} D. Goyeneche, 
Mutually unbiased triplets from non-affine families of complex Hadamard matrices in dimension 6, {\em J. Phys. A: Math. Theor.} 46 (2013) 105301 (15pp). 


			\bibitem{53Spengler} C. Spengler and B. Kraus, 
Graph-state formalism for mutually unbiased bases, {\em Phys. Rev. A} 88 (2013) 052323 (21pp). 


			\bibitem{54Olupitan} T. Olupitan, C. Lei and A. Vourdas, 
An analytic function approach to weak mutually unbiased bases, {\em Ann. Phys. (N.Y.)} 371 (2016) 1-19. 


			\bibitem{02Kibler} M.R. Kibler, 
Galois Fields and Galois Rings Made Easy, ISTE Press--Elsevier, London and Oxford (2017). 


			\bibitem{Sriwongsa} S. Sriwongsa and Y.M. Zou, 
Orthogonal abelian Cartan subalgebra decomposition of $sl_n$ over a finite commutative ring, {\em Linear and Multilinear Algebra} 
DOI: 10.1080/03081087.2018.1433626, arXiv:1802.02275 [math.RA] (2018). 


			\bibitem{Rao} H.S.S. Rao, S. Sirsi and K. Bharath, 
Mutually disjoint, maximally commuting set of physical observables for optimum state determination, arXiv:1809.06762 [quant-ph] (2018)


			\bibitem{Trifa55} Y. Trifa, 
Utilisation et construction de bases mutuellement non biais\'ees en th\'eorie de l’information quantique, {\em Rapport de stage, IPN Lyon -- ENS Lyon} (2018). 


			\bibitem{IBM56} https://quantumexperience.ng.bluemix.net/qx/experience (IBM Quantum Experience). 


			\bibitem{ATOS57} https://atos.net/en/insights-and-innovation/quantum-computing/atos-quantum (ATOS Quantum Learning Machine) 


}
\end{thebibliography}
\end{document}